\documentclass[journal]{IEEEtran}

\usepackage{amsmath}
\usepackage{amssymb}
\usepackage{amsthm}
\usepackage{epsfig}
\usepackage{epsf}
\usepackage{subfigure}
\usepackage{graphicx}
\usepackage{url}
\usepackage[]{authblk}
\usepackage{cite}

\def\BibTeX{{\rm B\kern-.05em{\sc i\kern-.025em b}\kern-.08em
    T\kern-.1667em\lower.7ex\hbox{E}\kern-.125emX}}

\newtheorem{corollary}{Corollary}
\newtheorem{theorem}{Theorem}
\newtheorem{definition}{Definition}

\newtheorem{lemma}{Lemma}

\newtheorem{remark}{Remark}

\newcounter{numcount}

\begin{document}

%%%%%%%%%%%%%%%%%%%%%%%%%%%%%%%%%%%%%%%%%%%%%%%%%%%
%%%%%%%%%%%%%%%%%%%%%%%%%%%%%%%%%%%%%%%%%%%%%%%%%%%
%%%%%%%%%%%%%%%%%%%%%%%%%%%%%%%%%%%%%%%%%%%%%%%%%%%

% \title{Capacity Region of the MISO Broadcast\\ Channel with Delayed CSIT to within\\ Constant Number of Bits}

\title{Approximate Capacity Region of the MISO Broadcast Channels with Delayed CSIT}

\author{Alireza~Vahid,
        Mohammad~Ali~Maddah-Ali,
        and~A.~Salman~Avestimehr
        \thanks{Alireza Vahid is with the School of Electrical and Computer Engineering, Duke University, Durham, NC, USA. Email: {\sffamily alireza.vahid@duke.edu}.}
        \thanks{A. Salman Avestimehr is with the Electrical Engineering Department, University of Southern California, Los Angeles, CA, USA. Email: {\sffamily avestimehr@ee.usc.edu}.}
\thanks{Mohammad~Ali~Maddah-Ali is with Bell Labs, Alcatel-Lucent, Holmdel, NJ, USA. Email: {\sffamily mohammadali.maddah-ali@alcatel-lucent.com}.}
\thanks{The work of A. S. Avestimehr and A. Vahid is in part supported by NSF Grants CAREER-0953117, CCF-1161720, NETS-1161904, AFOSR Young Investigator Program Award, ONR award N000141310094, and 2013 Qualcomm Innovation Fellowship.}
\thanks{The priliminary results of this work were presented at the Allerton Conference~\cite{MISOAllerton}.}
}
\maketitle

%%%%%%%%%%%%%%%%%%%%%%%%%%%%%%%%%%%%%%%%%%%%%%%%%%%
%%%%%%%%%%%%%%%%%%%%%%%%%%%%%%%%%%%%%%%%%%%%%%%%%%%
%%%%%%%%%%%%%%%%%%%%%%%%%%%%%%%%%%%%%%%%%%%%%%%%%%%

\begin{abstract}
We consider the problem of multiple-input single-output Broadcast Channels with Rayleigh fading where the transmitter has access to delayed knowledge of the channel state information. We first characterize the capacity region of this channel with two users to within constant number of bits for all values of the transmit power. The proposed signaling strategy utilizes the delayed knowledge of the channel state information and the previously transmitted signals, in order to create a signal of common interest for both receivers. This signal would be the quantized version of the summation of the previously transmitted signals. A challenge that arises in deriving the result for finite signal-to-noise ratio regimes is the correlation that exists between the quantization noise and the signal. To guarantee the independence of quantization noise and signal, we extend the framework of lattice quantizers with dither together with an interleaving step. For converse, we use the fact that the capacity region of this problem is upper-bounded by the capacity region of a physically degraded broadcast channel with no channel state information where one receiver has two antennas. Then, we derive an outer-bound on the capacity region of this degraded broadcast channel. Finally, we show how to extend our results to obtain the approximate capacity of the $K$-user multiple-input single-output Broadcast Channel with delayed knowledge of the channel state information at the transmitter to within $2 \log_2 \left( K + 2 \right)$ bits/s/Hz.
\end{abstract}


%%%%%%%%%%%%%%%%%%%%%%%%%%%%%%%%%%%%%%%%%%%%%%%%%%%
%%%%%%%%%%%%%%%%%%%%%%%%%%%%%%%%%%%%%%%%%%%%%%%%%%%
%%%%%%%%%%%%%%%%%%%%%%%%%%%%%%%%%%%%%%%%%%%%%%%%%%%

\section{Introduction}
\label{Section:Introduction}

In wireless networks, receivers estimate the channel state information (CSI) and pass this information to the transmitters through feedback mechanisms. The extent to which channel state information is available at the transmitters has a direct impact on the capacity of wireless networks and the optimal strategies. In fast-fading scenarios where the coherence time of the channel is smaller than the delay of the feedback channel, providing the transmitters with up-to-date CSI is practically infeasible. Consequently, we are left with no choice but to understand the behavior of wireless networks under such constraint.

Our objective is to understand the effect of lack of up-to-date CSI on the \emph{capacity} region of wireless networks by considering a fundamental building block, namely the multiple-input single-output (MISO) broadcast channel (BC). In the context of MISO BC, it has been shown that even completely stale CSIT (also known as delayed CSIT) can still be very useful and can change the scale of the capacity, measured by the degrees of freedom (DoF)~\cite{MAT_Delayed}. The degrees of freedom by definition provides a first order approximation of the capacity, thus it is mainly useful in understanding the behavior of the capacity in high power regimes. However, it is not a suitable measure for practical settings with finite signal-to-noise ratio (SNR). 

% The impact of delayed CSIT has also been studied in wireless networks with distributed transmitters. This includes studying the DoF region of multiple-antenna interference channels (ICs) and X channels with delayed CSIT~\cite{vaze2012degreesICMIMO,GhasemiX1,vaze2012degrees,Jafar_Retrospective,Abdoli2011IC-X-Arxiv}, the capacity region of two-user IC with binary fading with delayed CSIT~\cite{BFICAllerton, BFICISIT2012, vahid2013capacity}, the DoF region of $K$-user Gaussian IC and X channels with delayed CSIT~\cite{Jafar_Retrospective,Abdoli2011IC-X-Arxiv,javad2013layeredJ}, and multi-antenna two-user Gaussian IC with delayed CSIT and Shannon feedback~\cite{tandon2012degrees,vaze2011degrees}. Moreover, researchers have considered variations of the assumption on the available CSIT at the transmitter (see~\cite{GesbertCorrelated,de2013degrees,tandon2012synergistic,gou2012optimal,chen2013toward,chen2013symmetric}), these variations assume that in addition to the delayed CSIT, the transmitter has an imperfect estimate of the current channel realization.

We first consider the two-user MISO BC and we focus on the effect of delayed CSIT at finite SNR regimes as opposed to the asymptotic DoF analysis. While there is a strong body of work on broadcast channels with perfect channel state information (see~\cite{weingarten2004capacity,weingarten2006capacity,ozarow1984achievable}), no capacity result has been reported for the delayed CSIT scenario. There are some prior results in the literature (for example~\cite{GesbertPrecoding}) that have proposed and analyzed several achievability strategies at finite SNR regimes. Nonetheless, characterizing the capacity region of the two-user MISO BC with delayed CSIT has remained open. 

In this paper, we provide the first constant-gap approximation of the capacity region of the two-user MISO BC with delayed CSIT. We obtain an achievable scheme and an outer-bound on the capacity region, and we analytically show that they are within $1.81$ bits/sec/Hz per user, for all values of the transmit power. Our numerical analysis shows that the gap is in fact smaller and in the worst case, it is at most $1.1$ bits/sec/Hz per user. % By numerical analysis, we show that gap is in fact less than $0.5$ bits/sec/Hz per user.

The proposed achievability scheme for the two-user MISO BC with delayed CSIT has two phases. Phase~1 has two segments and in the first segment and the second segment, transmitter respectively sends messages intended for receivers one and two. In each one of these segments, the unintended receiver overhears and saves some signal (interference) which is only useful for the other receiver. At this point, transmitter can evaluate the overheard signals using the delayed CSI. In Phase~2, the transmitter will swap the overheard signals between the receivers by sending a signal of common interest. This signal is the quantized version of the summation of the previously transmitted signals. The swapping is performed by exploiting the overheard signals as available side-information at receivers side.  The overall information that each receiver collects in the two phases is enough to decode the intended message. Although the above two phases follows the scheme of~\cite{MAT_Delayed}, as we will show, some important additional ingredients are needed to make an approximately optimal scheme.

To derive the outer-bound, we create a physically degraded BC by providing the received signal of one user one to the other user. Then, since we know that feedback does not enlarge the capacity region of a physically degraded BC~\cite{ElGamal-Degraded}, we ignore the delayed knowledge of the channel state information at the transmitter (\emph{i.e.} no CSIT assumption). We then derive an outer-bound on the capacity region of this degraded broadcast channel which in turn serves as an outer-bound on the capacity region of the two-user MISO BC with delayed CSIT. We  show  that  the  achievable  rate  region  and  the  outer-bound are within $1.81$ bits/sec/Hz per user. Using numerical analysis, we can show that the gap is in fact smaller than $1.1$ bits/sec/Hz per user.

We then consider the $K$-user MISO BC with delayed CSIT and we focus on the symmetric capacity. We show how to extend our ideas for the achievability and converse to this setting, and we show that for the symmetric capacity the gap between the achievable rate and the outer-bound is less than $2 \log_2(K+2)$ bits/sec/Hz independent of the transmit power. We use this result to provide the approximate capacity of the $K$-user MISO BC with delayed CSIT.

% We show that the capacity region for this channel can be achieved by Gaussian input distribution. The capacity region for this degraded broadcast channel in turn provides an outer-bound on the capacity region of the two-user MISO complex Gaussian BC with delayed CSIT. To prove the optimality of Gaussian input distribution, we face several challenges which are due to the channel variations in time and the absence of instantaneous CSIT. We consider parallel copies of the ergodic degraded BC, and we use the ideas developed in~\cite{geng2012capacity} and~\cite{shomorony2012worst} to overcome these challenges and prove the optimality of Gaussian input distribution. Finally, we evaluate our outer-bound for Gaussian input distribution, and we  show  that  the  achievable  rate  region  and  the  outer-bound are within $1$ bit/sec/Hz per user.

% This work extends our earlier results in~\cite{MISOBCAllerton} by providing tighter bounds on the capacity region of the two-user MISO BC with delayed CSIT, and reducing the capacity approximation gap to $1$ bit/sec/Hz per user. As an intermediate step, we characterize the capacity region of the complex Gaussian multiple-input multiple-output (MIMO) degraded broadcast channel with no CSIT. The latter result along with some of the novel steps taken to prove it, could be of independent interest, and is the key step in deriving tighter bounds compared to~\cite{MISOBCAllerton}.

In the literature, there have been several results on the impact of delayed CSIT in wireless networks. However, these results are either focused on the DoF region (\emph{e.g},~\cite{Vaze_DCSIT_MIMO_BC,tandon2012degrees,chen2013toward}) or capacity results for noiseless channels (\emph{e.g.},~\cite{vahid2013capacity,vahid2013communication,vahid2015impact,LocalCSITJournal}). The present work provides constant-gap approximation of the capacity region of the two-user Complex Gaussian MISO BC with delayed CSIT which is of great importance in practical wireless settings. Existing results on constant-gap approximation of the capacity region of wireless networks mainly consider the scenario in which transmitters have perfect instantaneous knowledge of the channel state information (\emph{e.g},~\cite{Suh,AlirezaFB,Guy2}). Thus the current result opens the door for constant-gap approximation of the capacity region of wireless networks with delayed CSIT.

The rest of the paper is organized as follows. In Section~\ref{Section:Problem}, we formulate our problem. In Section~\ref{Section:Main}, we present our main results. We describe our achievability strategy in Section~\ref{Section:Achievability}. Section~\ref{Section:Converse} is dedicated to deriving the outer-bound. In Section~\ref{Section:Gap}, we show that our inner-bound and outer-bound are within constant number of bits. We extend our results to the $K$-user MISO BC with delayed CSIT in Section~\ref{Section:Kuser}. Finally, Section~\ref{Section:Conclusion} concludes the paper.

%%%%%%%%%%%%%%%%%%%%%%%%%%%%%%%%%%%%%%%%%%%%%%%%%%%
%%%%%%%%%%%%%%%%%%%%%%%%%%%%%%%%%%%%%%%%%%%%%%%%%%%
%%%%%%%%%%%%%%%%%%%%%%%%%%%%%%%%%%%%%%%%%%%%%%%%%%%

\section{Problem Setting}
\label{Section:Problem}

\begin{figure}[ht]
\centering
\subfigure[]{\includegraphics[height = 2.75cm]{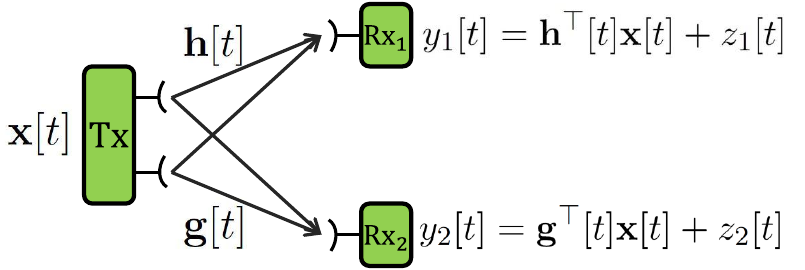}}
\subfigure[]{\includegraphics[height = 3.25cm]{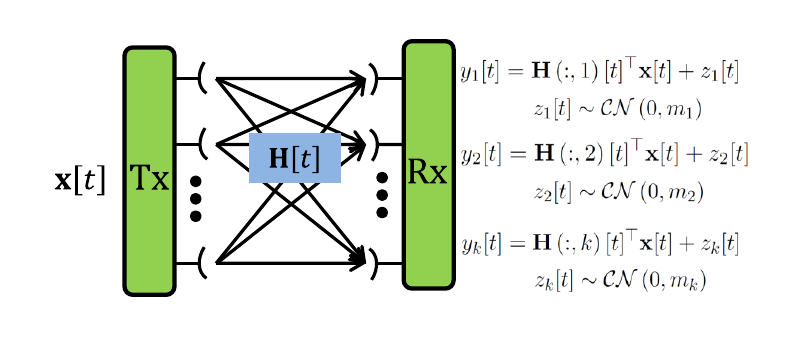}}
\caption{(a) Two-user Multiple-Input Single-Output (MISO) Complex Gaussian Broadcast Channel; (b) Point-to-point complex Gaussian channel with $k$ transmit and receive antennas and no CSIT, where $z_j[t] \sim \mathcal{CN}\left( 0, m_j \right)$. Here, $\mathbf{H}[t]$ denotes the channel transfer matrix at time $t$; and $\mathbf{H}\left( :,j\right)[t]$ denotes the channel gains from all transmit antennas to receive antenna $j$ at time $t$.}\label{Fig:MISO-BC-OFB}
\end{figure}

We start by considering the two-user multiple-input single-output (MISO) complex Gaussian broadcast channel (BC) with Rayleigh fading as depicted in Fig.~\ref{Fig:MISO-BC-OFB}(a). The channel gains from the transmitter to receivers one and two are denoted by $\mathbf{h}[t], \mathbf{g}[t] \in \mathbb{C}^{2 \times 1}$, respectively, where the entries of $\mathbf{h}[t]$ and $\mathbf{g}[t]$ are distributed as i.i.d. $\mathcal{CN}(0,1)$ (independent across time, antenna, and users). At each receiver, the received signal can be expressed as follows.
\begin{align}
y_1[t] = \mathbf{h}^\top[t] \mathbf{x}[t] + z_1[t], \qquad y_2[t] = \mathbf{g}^\top[t] \mathbf{x}[t] + z_2[t],
\end{align}
where $\mathbf{x}[t] \in \mathbb{C}^{2 \times 1}$ is the transmit signal subject to average power constraint $P$, \emph{i.e.} $\mathbb{E}\left[ \mathbf{x}^\dagger[t] \mathbf{x}[t] \right] \leq P$ for $P > 0$. The noise processes are independent from the transmit signal and are distributed i.i.d. as $z_k[t] \sim \mathcal{CN}(0,1)$. Furthermore, we define 
\begin{align}
\label{defineS}
s_1[t] = \mathbf{h}^\top[t] \mathbf{x}[t], \qquad s_2[t] = \mathbf{g}^\top[t] \mathbf{x}[t],
\end{align}
to be the noiseless versions of the received signals.

Transmitter wishes to reliably communicate independent and uniformly distributed messages $w_1 \in \{1,2,\ldots,2^{n R_1} \}$ and $w_2 \in \{1,2,\ldots,2^{n R_2} \}$ to receivers $1$ and $2$, respectively, during $n$ uses of the channel. We denote the channel state information at time $t$ by $\left( \mathbf{h}[t], \mathbf{g}[t] \right)$ for $t=1,2,\ldots,n$.
%\begin{align}
%\left( \mathbf{h}[t], \mathbf{g}[t] \right), \qquad t=1,2,\ldots,n.
%\end{align}
The transmitter has access to the delayed (outdated) channel state information, meaning that at time instant $t$, the transmitter has access to $\left( \mathbf{h}[\ell], \mathbf{g}[\ell] \right)_{\ell=1}^{t-1}$ for $t = 1,2,\ldots,n$.
%\begin{align}
%\left( \mathbf{h}[\ell], \mathbf{g}[\ell] \right)_{\ell=1}^{t-1}, \qquad t = 1,2,\ldots,n.
%\end{align}

Due to the delayed knowledge of the channel state information, the encoded signal $\mathbf{x}[t]$ is a function of both the messages and the previous channel realizations.

Each receiver $k$, $k=1,2$, uses a decoding function $\varphi_{k,n}$ to get the estimate $\hat{w}_k$ from the channel outputs $\{ y_{k}[t] : t = 1, \ldots, n \}$. An error occurs whenever $\hat{w}_k \neq w_k$. The average probability of error is given by $\lambda_{k,n} = \mathbb{E}[P(\hat{w}_k \neq w_k)]$, $k=1,2$,
%\begin{equation}\label{}
%\lambda_{k,n} = \mathbb{E}[P(\hat{w}_k \neq w_k)], \qquad k = 1, 2,
%\end{equation}
where the expectation is taken with respect to the random choice of the transmitted messages $w_1$ and $w_2$.

We say that a rate pair $(R_1,R_2)$ is achievable, if there exists a block encoder at the transmitter, and a block decoder at each receiver, such that $\lambda_{k,n}$ goes to zero as the block length $n$ goes to infinity, $k=1,2$. The capacity region $\mathcal{C}$ is the closure of the set of the achievable rate pairs.

%%%%%%%%%%%%%%%%%%%%%%%%%%%%%%%%%%%%%%%%%%%%%%%%%%%
%%%%%%%%%%%%%%%%%%%%%%%%%%%%%%%%%%%%%%%%%%%%%%%%%%%
%%%%%%%%%%%%%%%%%%%%%%%%%%%%%%%%%%%%%%%%%%%%%%%%%%%

\section{Statement of the Main Result}
\label{Section:Main}

Our main contributions are: $(1)$ characterization of the capacity region of the two-user MISO complex Gaussian BC with delayed CSIT to within $1.81$ bits/sec/Hz; and $(2)$ characterizing the symmetric capacity of the $K$-user MISO complex Gaussian BC to within $2 \log_2 \left( K + 2 \right)$ bits/sec/Hz. 

For the two-user setting, the achievability scheme has two phases. Phase~1 has two segments and in the first segment and the second segment, the transmitter respectively sends messages intended for receiver one and receiver two. In each of these segments, the unintended receiver overhears and saves some signal (interference), which is only useful for the other receiver. Later, in the second phase, the transmitter evaluates what each receiver overheard about the other receiver's message using the delayed knowledge of the channel state information and provides these overheard signals efficiently to both receivers exploiting available side information at each receiver. Transmitter provides the overheard signals to the receivers by sending a signal of common interest. This way transmitter reduces the overall communication time and in turn, increases the achievable rate.

The outer-bound is derived based on creating a physically degraded broadcast channel where one receiver is enhanced by having two antennas. In this channel, feedback and in particular delayed knowledge of the channel state information, does not increase the capacity region. Thus, we can ignore the delayed knowledge of the channel state information and consider a degraded BC with no CSIT. This would provide us with the outer-bound.

We then show how to extend our arguments for achievability and converse to the $K$-user setting to derive approximate symmetric capacity under delayed CSIT assumption. Before stating our results, we need to define some notations.

\begin{definition}
\label{def:tau}
For a region $\mathcal{R} \subseteq \mathbb{R}^2$, we define
\begin{align}
\mathcal{R} \ominus \left( \tau, \tau \right) \overset{\triangle}= \left\{ \left( R_1, R_2 \right) | R_1,R_2 \geq 0, \left( R_1 + \tau, R_2 + \tau \right) \in \mathcal{R} \right\}.
\end{align}
\end{definition}

\begin{definition}
\label{Def:Ckl}
Consider an ergodic point-to-point complex Gaussian channel with $k$ transmit antennas and $\ell$ receive antennas where the channel gains are distributed as i.i.d. $\mathcal{CN}(0,1)$ (independent across time, antenna, and users). At receive antenna $j$, we assume an additive zero-mean Gaussian noise process of variance $\sigma^2_j$. Then under no CSIT assumption and for a total average transmit power of $P$, the ergodic capacity of this channel is denoted by
\begin{align}
C_{k \times \ell}\left(P;\sigma^2_1,\sigma^2_2,\ldots,\sigma^2_{\ell} \right).
\end{align}
For simplicity of notations, we drop $P$, and whenever noise variances are all equal to $1$, we do not mention them. Fig.~\ref{Fig:MISO-BC-OFB}(B) depicts a point-to-point complex Gaussian channel with $k$ transmit antennas and $k$ receive antennas.
\end{definition}

%\begin{figure}[t]
%\centering
%\includegraphics[height = 4cm]{FiguresPDF/P2P-Kantenna.pdf}
%\caption{Point-to-point complex Gaussian channel with $k$ transmit and receive antennas and no CSIT, where $z_j[t] \sim \mathcal{CN}\left( 0, m_j \right)$. Here, $\mathbf{H}[t]$ denotes the channel transfer matrix at time $t$; and $\mathbf{H}\left( :,j\right)[t]$ denotes the channel gains from all transmit antennas to receive antenna $j$ at time $t$.}\label{Fig:P2P}
%\end{figure}

Consider the ergodic capacity of a point-to-point complex Gaussian channel with $2$ transmit antennas and a single receive antenna denoted by $C_{2 \times 1}\left( P;1 \right)$ as described in Definition~\ref{Def:Ckl}. Then from~\cite{Telatar-MIMO}, we have
\begin{align}
\label{Eq:P2P}
C_{2 \times 1}\left( P;1 \right) = C_{2 \times 1} = \mathbb{E}\log_2 \left[ 1 + \frac{P}{2} \mathbf{g}^\dagger \mathbf{g} \right],
\end{align}
where $\mathbf{g}$ is a $2$ by $1$ vector where entries are i.i.d. $\mathcal{CN}\left( 0, 1\right)$.

\begin{definition}
\label{Def:Ck}
Based on $C_{2 \times 1}$ described above, we define region $\mathcal{A}$ and region $\mathcal{B}$ as
\begin{align}
\label{eq:OuterBoundDelayedStar}
\mathcal{A} \overset{\triangle}= \left\{ \left( R_1, R_2 \right) \geq 0 \left| R_1, R_2  \geq 0, R_1 + 2 R_2 \leq 2 C_{2 \times 1} \right. \right\}, \nonumber \\
\mathcal{B} \overset{\triangle}= \left\{ \left( R_1, R_2 \right) \geq 0 \left| R_1, R_2  \geq 0, 2 R_1 + R_2 \leq 2 C_{2 \times 1} \right. \right\}.
\end{align}
\end{definition}
 
\begin{remark}
As we will show in Section~\ref{Section:Converse}, $\mathcal{A}$ ($\mathcal{B}$) is an outer-bound on the capacity region of a two-user complex Gaussian MIMO BC with no CSIT where ${\sf Rx}_1$ (${\sf Rx}_2$) has two antennas and ${\sf Rx}_2$ (${\sf Rx}_1$) has only one antenna (additive noise processes all having zero-mean and variance $1$). The corner points of $\mathcal{A}$ are given by $\left( 0, C_{2 \times 1} \right)$ and $\left( 2 C_{2 \times 1}, 0 \right)$.
\end{remark}

The following theorem states our contribution for the two-user MISO BC with delayed CSIT.

\begin{figure}[t]
\centering
\includegraphics[height = 6.5cm]{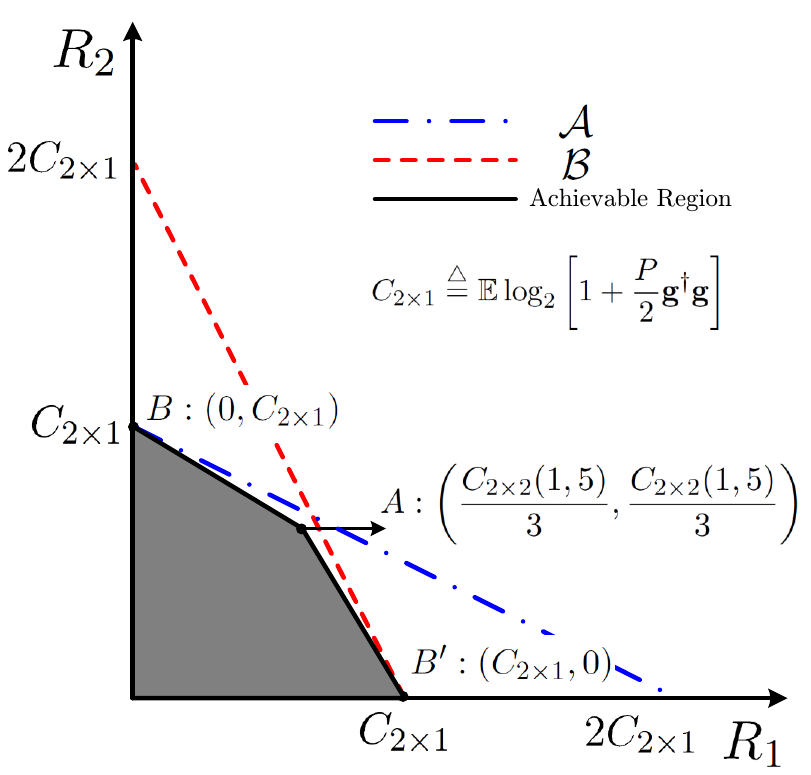}
\caption{The outer-bound on the capacity region of the two-user MISO BC with delayed CSIT is the intersection of $\mathcal{A}$ and $\mathcal{B}$. We prove that the capacity region is within $1.81$ bit/sec/Hz per user of this outer-bound. The achievable rate region is shown by the shaded area. $C_{2 \times 1}$ and $C_{2 \times 2}(1,5)$ are given in Definition~\ref{Def:Ckl}.}\label{Fig:RegionAll}
\end{figure}

\begin{theorem}
\label{THM:Main}
The capacity region of the two-user MISO BC with delayed CSIT, $\mathcal{C}$, is within $1.81$ bits/sec/Hz per user of $\left( \mathcal{A} \cap \mathcal{B} \right)$, \emph{i.e.}
\begin{align}
\left( \mathcal{A} \cap \mathcal{B} \right) \ominus \left( 1.81, 1.81 \right) \subseteq \mathcal{C} \subseteq \left( \mathcal{A} \cap \mathcal{B} \right), 
\end{align}
where $\mathcal{A}$ and $\mathcal{B}$ are given in Definition~\ref{Def:Ck}.
\end{theorem}

\begin{remark}
\label{remarkC22}
Fig.~\ref{Fig:RegionAll} pictorially depicts our result for the two-user MISO BC with delayed CSIT. We have defined $C_{2 \times 1}$ and $C_{2 \times 2}(1,5)$ in Definition~\ref{Def:Ckl}. We analytically show that the achievable region is within $1.81$ bits/sec/Hz per user of the outer-bound. Our numerical analysis shows that the gap is in fact smaller and in the worst case, it is at most $1.1$ bits/sec/Hz per user.
\end{remark}

We then consider the $K$-user MISO BC with delayed CSIT. We only focus on the symmetric capacity defined as follows.

\begin{definition}
\label{Def:SymK}
The symmetric capacity of the $K$-user MISO BC with delayed CSIT, $C_{\mathrm{SYM},\mathrm{K-user}}$, is given by
\begin{align}
& C_{\mathrm{SYM},\mathrm{K-user}} \overset{\triangle}= \sup \left\{ R:~R \leq R_j, \forall j = 1,2,\ldots,K, \right. \nonumber \\
& \left.~\left( R_1, \ldots, R_K \right) \in \right. \nonumber \\
& \left.~\text{capacity region of $K$-user MISO BC with delayed CSIT} \right\}.
\end{align}
\end{definition}

The following theorem states our contribution for the $K$-user setting.

\begin{theorem}
\label{THM:Kuser}
For the $K$-user MISO BC with delayed CSIT, $C_{\mathrm{SYM},\mathrm{K-user}}$ is lower-bounded by 
{\small \begin{align}
\frac{C_{K \times K}\left( 1, 5, \ldots, 1+(j-1)(j+2), \ldots, 1+(K-1)(K+2) \right)}{K\sum_{j=1}^{K}{j^{-1}}},
\end{align}}
and upper-bounded by $\left( {\sum_{j=1}^{K}{j^{-1}}} \right)^{-1}C_{K \times 1}$.
\end{theorem}

We use Theorem~\ref{THM:Kuser} to provide the approximate capacity of the $K$-user MISO BC with delayed CSIT in Section~\ref{Section:Kuser}.

\begin{remark}
For Theorem~\ref{THM:Kuser}, our numerical analysis shows that the gap is at most $2.3$ bits/sec/Hz per user for $K \leq 20$ and $P \leq 60 {\sf dB}$. In general, we show that the gap is at most $2\log_2(K + 2)$ bits/sec/Hz.
\end{remark}

%%%%%%%%%%%%%%%%%%%%%%%%%%%%%%%%%%%%%%%%%%%%%%%%%%%
%%%%%%%%%%%%%%%%%%%%%%%%%%%%%%%%%%%%%%%%%%%%%%%%%%%
%%%%%%%%%%%%%%%%%%%%%%%%%%%%%%%%%%%%%%%%%%%%%%%%%%%

\section{Achievability Proof of Theorem~\ref{THM:Main}}
\label{Section:Achievability}

In this section, we describe the achievability strategy of Theorem~\ref{THM:Main}. To characterize the capacity region of the two-user MISO complex Gaussian BC to within $1.81$ bits/sec/Hz per user, we need to show that the $\left( \mathcal{A} \cap \mathcal{B} \right)  \ominus \left( 1.81, 1.81 \right)$ is achievable.

We define rate region $\mathcal{R}(D)$ for parameter $D \ge 4$ as follows:
\begin{equation}
\label{eq:AchievableRegion}
\mathcal{R}(D) \overset{\triangle}=
\left\{ \begin{array}{ll}
\left( R_1, R_2 \right) \left| \parbox[c][4em][c]{0.2\textwidth} {$0 \leq R_1 + \gamma R_2 \leq C_{2 \times 1}$ \\
$0 \leq \gamma R_1 + R_2 \leq C_{2 \times 1}$} \right.
\end{array} \right\}
\end{equation}
where $\gamma = \left( \frac{3 C_{2 \times 1}}{C_{2 \times 2}(1,1+D) } - 1 \right)$\footnote{As we will see later in this section and Section~\ref{Section:Gap}, we have $\left( \frac{3 C_{2 \times 1}}{C_{2 \times 2}(1,1+D) } - 1 \right) > 0$ for $D \ge 4$ and $P > 0 {\sf dB}$.}.
 
Later in Section~\ref{Section:Gap}, we show that $\left( \mathcal{A} \cap \mathcal{B} \right)  \ominus \left( 1.81, 1.81 \right) \subseteq \mathcal{R}(4)$, and thus, to characterize the capacity region to within $1.81$ bit/sec/Hz per user, it suffices to show that $\mathcal{R}(4)$ is achievable. 

The shadowed region in Fig.~\ref{Fig:RegionAll} corresponds to $\mathcal{R}(4)$, which is a polygon with corner points $A,B,$ and $B^\prime$. Corner points $B$ and $B^\prime$ of $\mathcal{R}(4)$, are achievable using the result on point-to-point MISO Gaussian channels with no CSIT~\cite{Telatar-MIMO}. Therefore, we only need to describe the achievability strategy for corner point $A$.

\subsection{Transmission Strategy for Corner Point $A$}
\label{Section:AchievabilityStrategy}

Here, we present the achievability strategy for corner point $A$. We show that for any $\epsilon > 0$, we can achieve
\begin{align} 
\left( R_1, R_2 \right) = \left( \frac{C_{2 \times 2}(1,5) - \epsilon}{3}, \frac{C_{2 \times 2}(1,5) - \epsilon}{3} \right),
\end{align}
with vanishing decoding error probability as the communication length goes to infinity.

The achievability strategy is carried on over $n$ blocks, each block consisting of $2$ phases. Phase $1$ has two segments each of length $n$ channel uses, and Phase $2$ is of length $n$ channel uses. Denote by $w_k^b$, the message of user $k$ in block $b$, $k \in \{ 1,2 \}$, $b = 1,2,\ldots,n$. Fix $\epsilon > 0$ and set $R = C_{2 \times 2}(1,1+D) - \epsilon$. We assume that $w_k^b \in \{ 1,2,\ldots,2^{nR} \}$ and that the messages are distributed uniformly and independently. The encoding is carried on as described below.

$\bullet$ {\bf Encoding:} At the transmitter, the message of user $k$ during block $b$, \emph{i.e.} $w^b_k$, is mapped to a codeword of length $n$ denoted by $\mathbf{x}_k^{b,n}$ where any element of this codeword is drawn i.i.d. from $\mathcal{CN} \left( 0, P/2 \mathbf{I}_2 \right)$\footnote{$\mathbf{I}_2$ is the $2 \times 2$ identity matrix.}.

$\bullet$ Communication during segment $j$ of Phase $1$ of block $b$: During this segment, the transmitter communicates $\mathbf{x}_j^{b,n}$ from its two transmit antennas, $j = 1,2$, and $b = 1,2,\ldots,n$. Receiver one obtains $y_{1,1:j}^{b,n}$ and receiver two obtains $y_{2,1:j}^{b,n}$.

\vspace{1mm}

$\bullet$ Communication during Phase $2$ of block $b$: Using the delayed CSIT, the transmitter creates
\begin{align}
\label{eq:S}
s^{b,n} = s_{2,1:1}^{b,n} + s_{1,1:2}^{b,n},
\end{align}
where $s_{2,1:1}^{b,n}$ is the received signal at ${\sf Rx}_2$ during the first segment of Phase $1$ of block $b$, $y_{2,1:1}^{b,n}$, minus the noise term as defined in (\ref{defineS}), and $s_{1,1:2}^{b,n}$ is the received signal at ${\sf Rx}_1$ during the second segment of Phase $1$ of block $b$, $y_{1,1:2}^{b,n}$, minus the noise term.

Note that $s_{2,1:1}^{b,n} + s_{1,1:2}^{b,n}$ is useful for both receivers since each receiver can subtract its previously received signal to obtain what the other receiver has (up to the noise term). Therefore, the goal in this phase, is to provide $s_{2,1:1}^{b,n} + s_{1,1:2}^{b,n}$ to both receivers with distortion $D = 4$. 

\begin{remark}
We note that the idea of creating quantized version of previously received signals has been utilized previously for asymptotic degrees of freedom analysis~\cite{kobayashi2012degrees,chen2012degrees,chen2014vector}. However for finite SNR regime, we need to take into account the fact that the quantization noise is neither independently distributed over time nor is it independent from the signal; and we need to overcome these challenges. 
\end{remark}

\begin{figure}[ht]
\centering
\includegraphics[height = 3.5cm]{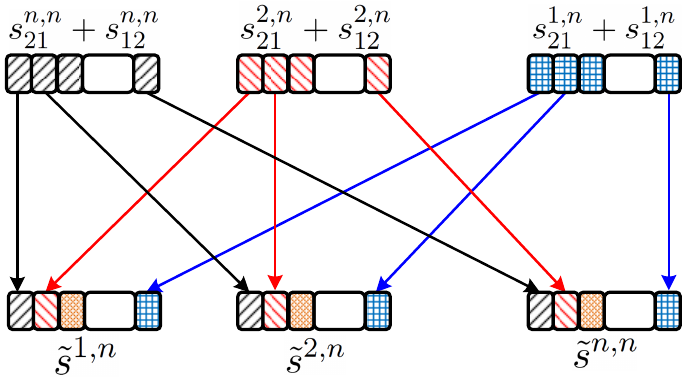}
\caption{The interleaving step: the resulting signal at different time instants of a given phase at a given block are independent from each other.}\label{Fig:Interleaving}
\vspace{-5mm}
\end{figure}

The input signal to a lattice quantizer needs to be independently distributed over time (see~\cite{Zamir_Lattice_Noise,Zamir_Nested}). Thus, in order to quantize this signal using lattice quantizer, we need  
$$\left( s_{2,1:1}^{b}[\ell] + s_{1,1:2}^{b}[\ell] \right)_{\ell=1}^n$$
to be an independently distributed sequence. However, given message $w^b_k$, the transmit signal $\mathbf{x}_k^{b,n}$ is correlated across time and as a result, the aforementioned signals are not independent anymore. In order to overcome this issue, we incorporate an interleaving step according to the following mapping which is depicted in Fig.~\ref{Fig:Interleaving}.
\begin{align}
\label{eq:S1}
\tilde{s}^b[t] = s_{2,1:1}^{t}[b] + s_{1,1:2}^{t}[b],
\end{align}
where $b = 1,2,\ldots,n$ and $t = 1,2,\ldots,n$. It is important to notice that with this interleaving, the resulting signal at different time instants of a given phase at a given block are independent from each other. This is due to the fact that these signals are created from independent messages.

Note that given the previous channel realizations, the signal in (\ref{eq:S1}) at each time has a Gaussian distribution but its variance varies from each time instant to the other. Thus, in order to be able to quantize it, we need a generalization of the rate-distortion function to include non-identically distributed sources. Below, we discuss this issue.

\begin{lemma}
\label{Lemma:RateDistortion}
{[Rate distortion for non-identically distributed Gaussian Source]}~Consider a independently non-identically distributed Gaussian source $\mathbf{u}$, where at time instant $t$, it has zero mean and variance $\sigma^2[t]$. Assume that $\sigma^2[t]$ is drawn from some i.i.d. continuous distribution with $\mathbb{E} \left[ \sigma^2 \right] < \infty$. The sequence of $\sigma^2[t]$ is non-causally known by both encoder and decoder. Then, with squared-error distortion, we can quantize the signal at any rate greater than or equal to
\begin{align}
\label{eq:ratedistortion}
\min_{D_\sigma: \mathbb{E}\left[ D_\sigma \right] \leq D}{\mathbb{E} \left[ \log_2 \frac{\sigma^2[t]}{D_\sigma} \right]^+},
\end{align}
and achieve distortion $D$ (per sample), where the expectation is with respect to the distribution of $\sigma^2$.
\end{lemma}

{\it Proof sketch:} Suppose $\sigma^2$ could only take $m$ finite values $\sigma_i^2$ with probability $p_i$, $i=1,2,\ldots,m$. The problem would then be similar to that of $m$ parallel Gaussian channels where waterfilling gives the optimal solution ((Theorem~10.3.3~\cite{cover2012elements})). For the time instants where source has variance $\sigma_i^2$, we choose a distortion $D_i$ such that $\sum_{i=1}^{m} p_i D_i \leq D$. Note that in order to derive the optimum answer, we need to optimize over the choice of $D_i$'s. The case where $\sigma^2$ take values in a continuous set can be viewed as the limit of the discrete scenario using standard arguments. \hfill{$\blacksquare$}

It is easy to see that any rate greater than or equal to
\begin{align}
\label{eq:ratedistortionsuboptimal}
\mathbb{E} \left[ \log_2 \left( 1 + \frac{\sigma^2[t]}{D} \right) \right],
\end{align}
is also achievable at distortion $D$ (per sample). Basically, we have ignored the optimization over $D$ and added a $1$ to remove $\max \{ ., 0\}$ (or $.^+$).  

Moreover, we would like the distortion to be independent of the signals. In order to have a distortion that is independent of the signal and is uncorrelated across time, we can incorporate lattice quantization with ``dither'' as described in~\cite{Zamir_Lattice_Noise}. Basically, dither is a random vector distributed uniformly over the basic Voronoi that is added to the signal before feeding it to the quantizer. At the decoder, this random vector will be subtracted.

From (\ref{eq:ratedistortionsuboptimal}), we conclude that we can quantize $\tilde{s}^{b,n}$ with squared-error distortion $D$ at rate $R_Q(D)$ (per sample), defined as
\begin{align}
\label{eq:ratedistortionsuboptimalconcave}
R_Q(D) \overset{\triangle}= \mathbb{E}\left[ \log_2 \left( 1 + \frac{P}{2D} \left( ||\mathbf{g}||_2^2 + ||\mathbf{h}||_2^2 \right) \right)\right],
\end{align}
where $\mathbf{g}, \mathbf{h} \in \mathbb{C}^{2 \times 1}$ with i.i.d. $\mathcal{CN}(0,1)$ entries.

We can reliably communicate the quantization index over 
\begin{align}
\left\lceil \frac{n R_Q(4)}{C_{2 \times 1}} \right\rceil 
\end{align}
time instants. Next, we need to show that given the appropriate choice of parameters, receivers can recover the corresponding messages with vanishing error probability as $n \rightarrow \infty$.

\begin{figure}[h]
\centering
\includegraphics[width = \columnwidth]{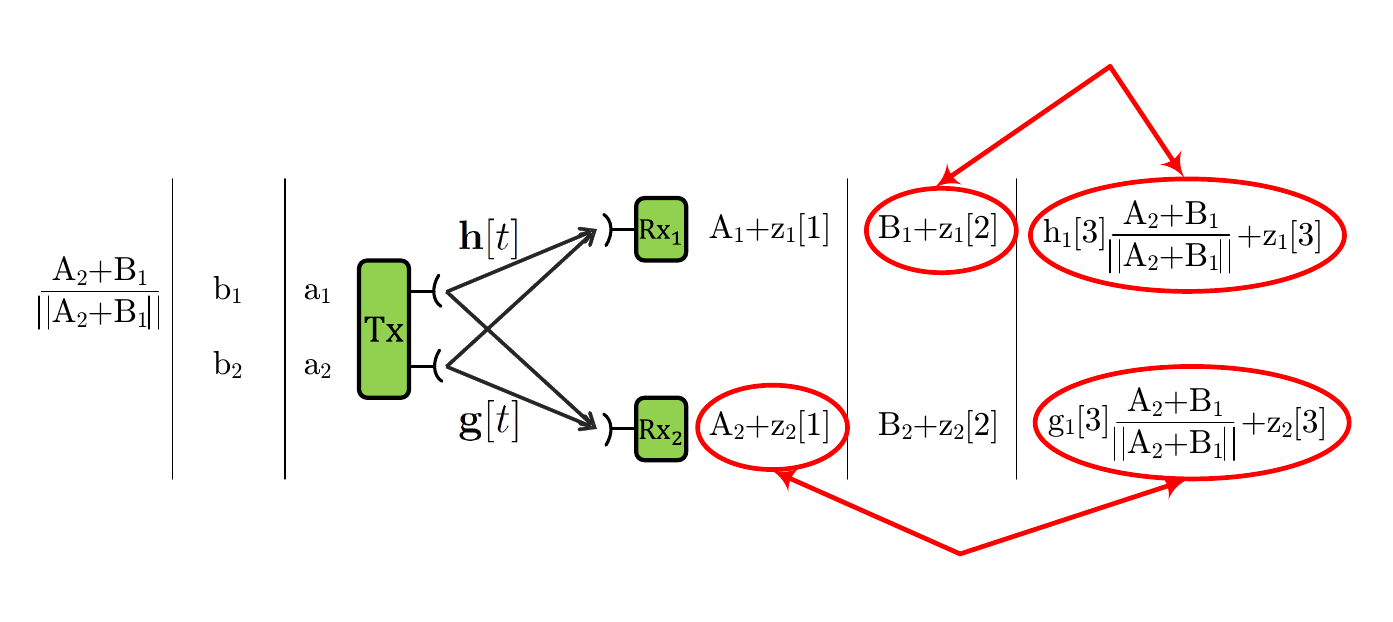}
\caption{The original three-slot Maddah-Ali-Tse protocol.}\label{Fig:MAT}
\end{figure}

\begin{remark}
An attempt to directly implement Maddah-Ali-Tse protocol in the finite SNR regime faces some challenges. Maddah-Ali-Tse protocol is depicted in Fig.~\ref{Fig:MAT}. At time 1, symbols $a_1$ and $a_2$ intended for receiver $1$ are transmitted and receiver $j$ obtains $A_j+z_j[1]$ where $z_j[1]$ is the additive noise and $A_j$ is result of multiplying the channel vector to receiver $j$ with the transmit signal. Similar story holds for time 2 where symbols $b_1$ and $b_2$ intended for receiver $2$ are transmitted. During time 3, $\frac{A_2+B_1}{||A_2+B_1||}$ is transmitted from the first antenna. For DoF purposes where we can ignore the noise terms, this scheme works by providing each receiver with sufficient information to decode its symbols.

Suppose due to the channel realization, $A_2$ has a much higher power compared to $B_1$, \emph{i.e.} $||A_2|| \gg ||B_1||$. The signal in time 3 is normalized to meet the power constraint and then transmitted. When receiver 2 removes $A_2$, it is left with $B_1+\tilde{z}_2[3]$ and the noise term may very well have a much higher power than $B_1$. Therefore, $B_1$ is effectively eliminated and this results in a significant rate loss and thus a bigger gap. 

To overcome this challenge, we incorporate quantization and allow the entire useful signal for each receiver to be quantized individually and then we create the XOR of the quantized bits as our transmit signal. This way we look at many samples similar to $A_2$ and $B_1$, and we effectively take into account the average power of such signals. In this manner, no signal will be eliminated due to have a lower power.
\end{remark}

\subsection{Decoding}

Upon completion of Phase $2$ of block $b$, each receiver decodes the quantized signal. We know that as $n \rightarrow \infty$, this could be done with arbitrary small decoding error probability. Therefore, each receiver has access to  $\tilde{s}^{b,n} + z_Q^{b,n}$
%\begin{align}
%\tilde{s}^{b,n} + z_Q^{b,n},
%\end{align}
where $z_Q^{b,n}$ is the quantization noise with variance $D$ which is independent of the transmit signals. Note that $z_Q^b[t_1]$ and $z_Q^b[t_2]$ are uncorrelated but not necessarily independent, $t_1,t_2 = 1,2,\ldots,n$, $t_1 \neq t_2$.

Receiver $1$ at the end of the $n^{th}$ communication block, reconstructs signals by reversing the interleaving procedure described above, and removes $y_{1,2}^{b,n}$ to obtain 
\begin{align}
\tilde{y}_{2,1:1}^{b,n} = y_{2,1:1}^{b,n} + \tilde{z}_Q^{b,n},
\end{align}
here $\tilde{z}_Q^{b,n}$ is the quantization noise with variance $D$ which is independent of the transmit signals. Moreover, $\tilde{z}_Q^b[\ell]_{\ell=1}^n$ is an \emph{independent} sequence.

Note that since the messages are encoded at rate $C_{2 \times 2}(1,1+D) - \epsilon$ for $\epsilon > 0$, if receiver one has access to $y_{2,1:1}^{b,n}$ up to distortion $D$, it can recover $w^b_1$ with arbitrary small decoding error probability as $\epsilon \rightarrow 0$ and communication length goes to infinity. Thus, from $y_{1,1:1}^{b,n}$ and $\tilde{y}_{2,1:1}^{b,n}$, receiver one can decode $w^b_1$, $b=1,2,\ldots,n$. Similar argument holds for receiver two.

An error may occur in either of the following steps: $(1)$ if an error occurs in decoding message $w^b_k$ provided required signals to the receiver, $k=1,2$; $(2)$ if an error occurs in quantizing $\tilde{s}^{b,n}$; and $(3)$ if an error occurs in decoding $\tilde{s}^{b,n} + z_Q^{b,n}$ at either of the receivers, $b=1,2,\ldots,n$. The probability of each one of such errors decreases exponentially in $n$ (see~\cite{Gersho,Telatar-MIMO} and references therein). Using union bound and given that we have $O\left( n^2 \right)$ possible errors and the fact that each error probability decreases exponentially to zero, the total error probability goes to zero as $n \rightarrow \infty$.

\subsection{Achievable Rate}

Using the achievable strategy described above, as $n \rightarrow \infty$, we can achieve a (symmetric) sum-rate point of
\begin{align}
\left( R_1, R_2 \right) = \left( \frac{C_{2 \times 2}(1,5)}{ 2 + R_Q(4)/C_{2 \times 1}}, \frac{C_{2 \times 2}(1,5)}{ 2 + R_Q(4)/C_{2 \times 1}} \right).
\end{align}

In Appendix~\ref{Appendix:ValD}, we show that $R_Q(4)/C_{2 \times 1} \leq 1$ for all values of $P$. Therefore, Phase $2$ of each block has at most $n$ time instants. Thus, we conclude that a (symmetric) sum-rate point of
\begin{align}
\label{eq:achrate}
\left( R_1, R_2 \right) = \left( \frac{C_{2 \times 2}(1,5)}{3}, \frac{C_{2 \times 2}(1,5)}{3} \right),
\end{align}
is achievable.

%%%%%%%%%%%%%%%%%%%%%%%%%%%%%%%%%%%%%%%%%%%%%%%%%%%
%%%%%%%%%%%%%%%%%%%%%%%%%%%%%%%%%%%%%%%%%%%%%%%%%%%
%%%%%%%%%%%%%%%%%%%%%%%%%%%%%%%%%%%%%%%%%%%%%%%%%%%

\section{Converse Proof of Theorem~\ref{THM:Main}}
\label{Section:Converse}

In this section, we provide the converse proof of Theorem~\ref{THM:Main}. The converse consists of two main parts. In part $1$, we show that the capacity region of the problem is included in the capacity region of a (stochastically) degraded BC, and in part $2$, we derive an outer-bound on the capacity region of the degraded BC.

\begin{figure}[ht]
\centering
\includegraphics[height = 4cm]{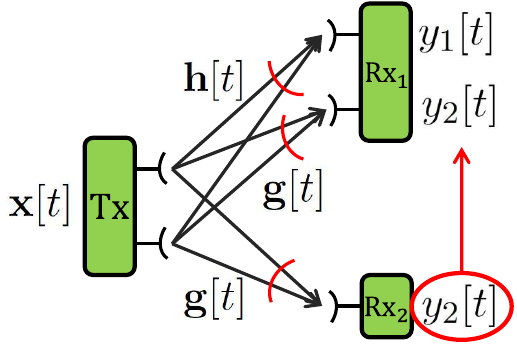}
\caption{By providing $y_2^n$ to ${\sf Rx}_1$, we create a physically degraded BC. We then ignore the delayed knowledge of the channel state information at the transmitter because for a physically degraded broadcast channel, feedback does not enlarge the capacity region. The resulting channel is stochastically degraded.}\label{Fig:MISO-BC-Degraded}
\end{figure}

\noindent {\bf Part 1:} We create the stochastically degraded BC as follows. We first provide the received signal of ${\sf Rx}_2$, \emph{i.e.} $y_2[t]$, to ${\sf Rx}_1$ as depicted in Fig.~\ref{Fig:MISO-BC-Degraded}. Note that the resulting channel is \emph{physically degraded}, and we know that for a physically degraded broadcast channel, feedback does not enlarge the capacity region~\cite{ElGamal-Degraded}. Therefore, we can ignore the delayed knowledge of the channel state information at the transmitter (\emph{i.e.} no CSIT assumption) and the correlation between the channel gains. The resulting channel is stochastically degraded.

Thus, we form a \emph{stochastically degraded} channel as shown in Fig.~\ref{Fig:MISO-BC-Degraded}, and  we formally define it in Definition~\ref{Def:MIMOBC} below. Let us denote an outer-bound on the capacity region of this channel by $\mathcal{A}$. The argument above shows that $\mathcal{C} \subseteq \mathcal{A}$.

\begin{definition}
\label{Def:MIMOBC}
The input-output relationship of a two-user stochastically degraded MIMO BC as depicted in Fig.~\ref{Fig:MISO-BC-Degraded}, is given by
\begin{align}
\label{eq:channelinput}
\tilde{\mathbf{y}}_1[t] = \mathbf{H}[t] \mathbf{x}[t] + \mathbf{z}_1[t], \qquad \tilde{y}_2[t] = \mathbf{g}^{\top}[t] \mathbf{x}[t] + z_2[t],
\end{align}
for
\begin{align}
\label{eq:H} 
\mathbf{H}[t] = 
\left[ \begin{array}{cc}
h_1[t] & h_2[t] \\
g_1[t] & g_2[t] \end{array} \right],
\qquad \mathbf{g}[t] = 
\left[ \begin{array}{c}
g_1[t] \\
g_2[t] \end{array} \right],
\end{align}
where $h_j[t]$ and $g_j[t]$ are distributed as i.i.d. (independent over time and from each other) $\mathcal{CN}(0,1)$ as described in Section~\ref{Section:Problem} for $j=1,2$. We have
\begin{align}
\tilde{\mathbf{y}}_1[t] = \left[~\tilde{y}_{11}[t] \quad \tilde{y}_{12}[t]~\right]^\top.
%\left[ \begin{array}{c}
%\tilde{y}_{11}[t] \\
%\tilde{y}_{12}[t] \end{array} \right].
\end{align}
We assume $\mathbf{z}_1[t] \in \mathbb{C}^{2 \times 1}$ and $z_2[t] \sim \mathcal{CN}(0,1)$, and the transmit signal is subject to average power constraint $P$. We further assume that the transmitter has no knowledge of the channel gains besides their distributions (no CSIT assumption).
\end{definition}

% \footnote{In this section, the transfer matrix from the transmitter to ${\sf Rx}_1$ is denoted by $\mathbf{H}$ which is a $2 \times 2$ matrix.}

\noindent {\bf Part 2:} In this part, we derive an outer-bound, $\mathcal{A}$, on the capacity region of the stochastically degraded broadcast channel. For the stochastically degraded BC defined in Definition~\ref{Def:MIMOBC}, suppose there exists encoders and decoders at the transmitter and receivers such that each message can be decoded at its corresponding receiver with arbitrary small decoding error probability. 
\begin{align}
\label{eq:boundR12R2}
n & \left( R_1 + 2 R_2 - 3 \epsilon_n \right) \nonumber \\
& \overset{\mathrm{Fano}}\leq I\left( w_1 ; \tilde{y}_{11}^n, \tilde{y}_{12}^n | w_2, \mathbf{H}^n, \mathbf{g}^n \right) + 2 I\left( w_2 ; \tilde{y}_{2}^n | \mathbf{H}^n, \mathbf{g}^n \right) \nonumber \\
& \overset{(a)}= I\left( w_1 ; \tilde{y}_{11}^n, \tilde{y}_{12}^n | w_2, \mathbf{H}^n, \mathbf{g}^n \right) + I\left( w_2 ; \tilde{y}_{11}^n | \mathbf{H}^n, \mathbf{g}^n \right) \nonumber \\
& \quad + I\left( w_2 ; \tilde{y}_{12}^n | \mathbf{H}^n, \mathbf{g}^n \right) \nonumber \\
& = h\left( \tilde{y}_{11}^n, \tilde{y}_{12}^n | w_2, \mathbf{H}^n, \mathbf{g}^n \right) - h\left( \tilde{y}_{11}^n, \tilde{y}_{12}^n | w_1, w_2, \mathbf{H}^n, \mathbf{g}^n \right) \nonumber \\
& \quad + h\left( \tilde{y}_{11}^n | \mathbf{H}^n, \mathbf{g}^n \right) - h\left( \tilde{y}_{11}^n | w_2, \mathbf{H}^n, \mathbf{g}^n \right) \nonumber \\
& \quad + h\left( \tilde{y}_{12}^n | \mathbf{H}^n, \mathbf{g}^n \right) - h\left( \tilde{y}_{12}^n | w_2, \mathbf{H}^n, \mathbf{g}^n \right) \nonumber \\
& \overset{(b)}= h\left( \tilde{y}_{11}^n | \mathbf{H}^n, \mathbf{g}^n \right) - h\left( z_{11}^n | \mathbf{H}^n, \mathbf{g}^n \right) \nonumber \\
& \quad + h\left( \tilde{y}_{12}^n | \mathbf{H}^n, \mathbf{g}^n \right) - h\left( z_{12}^n | \mathbf{H}^n, \mathbf{g}^n \right) \nonumber \\
& \quad + h\left( \tilde{y}_{11}^n, \tilde{y}_{12}^n | w_2, \mathbf{H}^n, \mathbf{g}^n \right) - h\left( \tilde{y}_{11}^n | w_2, \mathbf{H}^n, \mathbf{g}^n \right) \nonumber \\
& \quad - h\left( \tilde{y}_{12}^n | w_2, \mathbf{H}^n, \mathbf{g}^n \right) \nonumber \\
& \overset{(c)}\leq 2 \mathbb{E}\log_2 \left[ 1 + \frac{P}{2} \mathbf{g}^\dagger \mathbf{g} \right] - I\left( \tilde{y}_{11}^n; \tilde{y}_{12}^n| w_2, \mathbf{H}^n, \mathbf{g}^n \right) \nonumber \\
& \overset{(d)}\leq 2 \mathbb{E}\log_2 \left[ 1 + \frac{P}{2} \mathbf{g}^\dagger \mathbf{g} \right] \overset{(\ref{Eq:P2P})}= 2 C_{2 \times 1},
\end{align}
% Fano's inequality, the mutual independence of the messages and the channel realizations, and 
where $(a)$ follows from the fact that due to no CSIT assumption, we have 
\begin{align}
& h\left( w_2 |  \tilde{y}_{11}^n, \mathbf{H}^n, \mathbf{g}^n \right) \leq n \epsilon_n, \qquad h\left( w_2 |  \tilde{y}_{12}^n, \mathbf{H}^n, \mathbf{g}^n \right) \leq n \epsilon_n;
\end{align}
$(b)$ holds since
\begin{align}
& h\left( \tilde{\mathbf{y}}_1^n | w_1, w_2, \mathbf{H}^n, \mathbf{g}^n \right) = h\left( \tilde{\mathbf{y}}_1^n | w_1, w_2, \mathbf{x}^n, \mathbf{H}^n, \mathbf{g}^n \right) \nonumber \\
& \quad = h\left( z_{11}^n, z_{12}^n | w_1, w_2, \mathbf{x}^n, \mathbf{H}^n, \mathbf{g}^n \right) \nonumber \\
& \quad = h\left( z_{11}^n| \mathbf{H}^n, \mathbf{g}^n \right) + h\left( z_{12}^n| \mathbf{H}^n, \mathbf{g}^n \right);
\end{align}
$(c)$ follows from the results in~\cite{Telatar-MIMO}; and $(d)$ follows from fact that mutual information is always positive. Dividing both sides by $n$ and letting $n \rightarrow \infty$, we obtain the desired result. This completes the derivation of $\mathcal{A}$. Similarly, we can derive $\mathcal{B}$, and we have $\mathcal{C} \subseteq \mathcal{A} \cap \mathcal{B}$ which completes the converse proof for Theorem~\ref{THM:Main}.

%%%%%%%%%%%%%%%%%%%%%%%%%%%%%%%%%%%%%%%%%%%%%%%%%%%
%%%%%%%%%%%%%%%%%%%%%%%%%%%%%%%%%%%%%%%%%%%%%%%%%%%
%%%%%%%%%%%%%%%%%%%%%%%%%%%%%%%%%%%%%%%%%%%%%%%%%%%

\section{Gap Analysis}
\label{Section:Gap}

In this section, we evaluate the gap between our achievable rate-region and the outer-bound. We analytically prove that the gap is at most $1.81$ bits/sec/Hz per user. We show that
\begin{align}
\left( \mathcal{A} \cap \mathcal{B} \right) \ominus \left( 1.81, 1.81 \right) \subseteq \mathcal{R}(4),
\end{align}
where $\mathcal{A}$ and $\mathcal{B}$ are given in Definition~\ref{Def:Ck}, and $\mathcal{R}(4)$ is given in (\ref{eq:AchievableRegion}).

Since the achievable rate region and the outer-bound (see Fig.~\ref{Fig:RegionAll}) are formed by time sharing among the corresponding corner points (and thus, characterized by straight lines), we only need to consider the symmetric capacity, $\mathcal{C}^{2}_{\mathrm{SYM}}$, defined in Definition~\ref{Def:SymK}.

%\begin{definition}
%The symmetric capacity, $\mathcal{C}_{\mathrm{SYM}}$, is given by
%\begin{align}
%\mathcal{C}_{\mathrm{SYM}} = \sup \left\{ R: \left( R, R \right) \in \mathcal{C} \right\}.
%\end{align}
%\end{definition}

We evaluate the gap between the inner-bound in (\ref{eq:achrate}), \emph{i.e.} 
\begin{align}
\left( R_1, R_2 \right) = \left( \frac{C_{2 \times 2}(1,5)}{3}, \frac{C_{2 \times 2}(1,5)}{3} \right).
\end{align}
and the symmetric point $\left( \mathcal{C}_{\mathrm{SYM},\mathrm{2-user}}, \mathcal{C}_{\mathrm{SYM},\mathrm{2-user}} \right)$, obtained from Theorem~\ref{THM:Main}.

\begin{figure}[ht]
\centering
\subfigure[]{\includegraphics[height = 4cm]{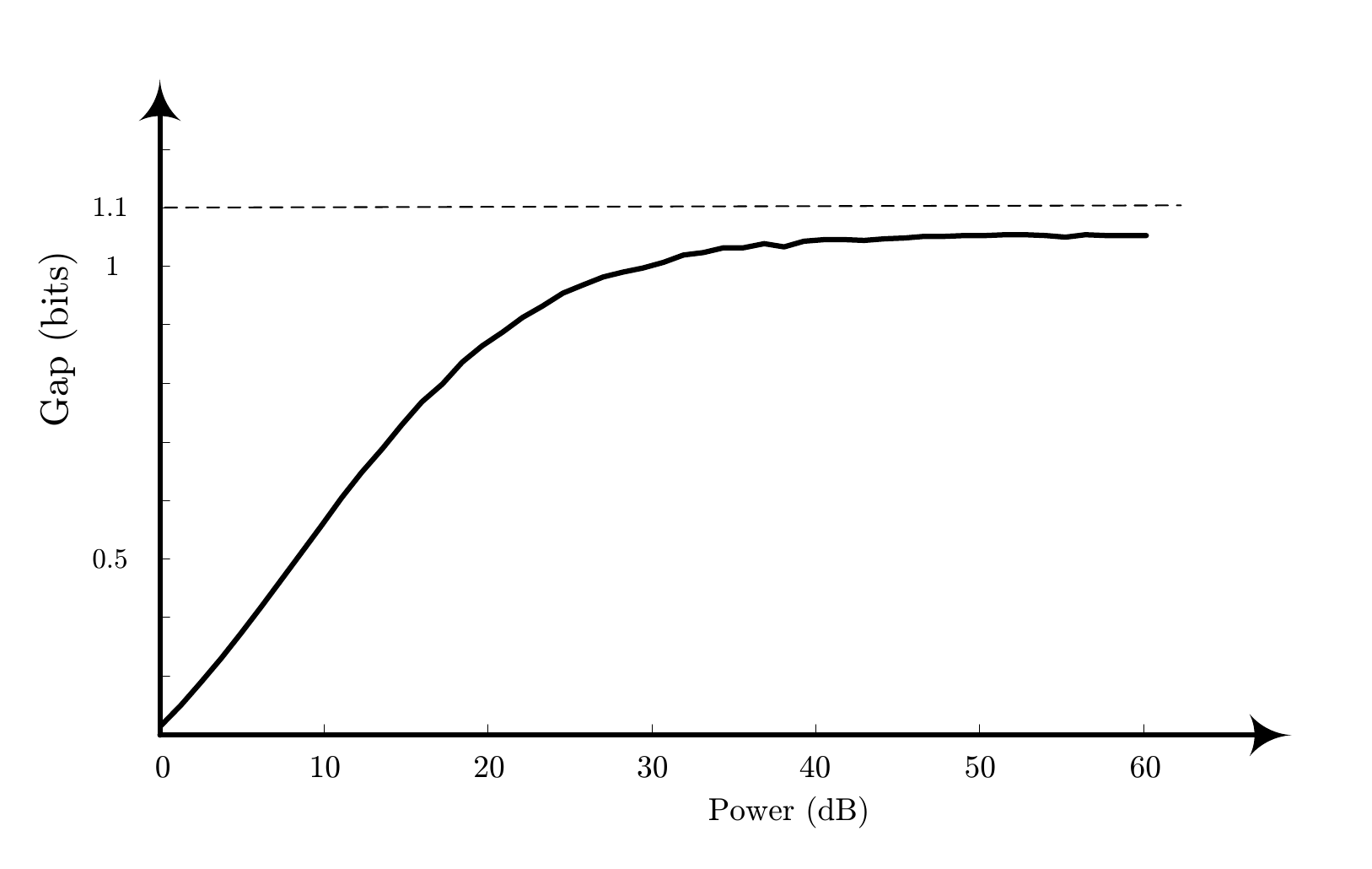}}
\subfigure[]{\includegraphics[height = 4cm]{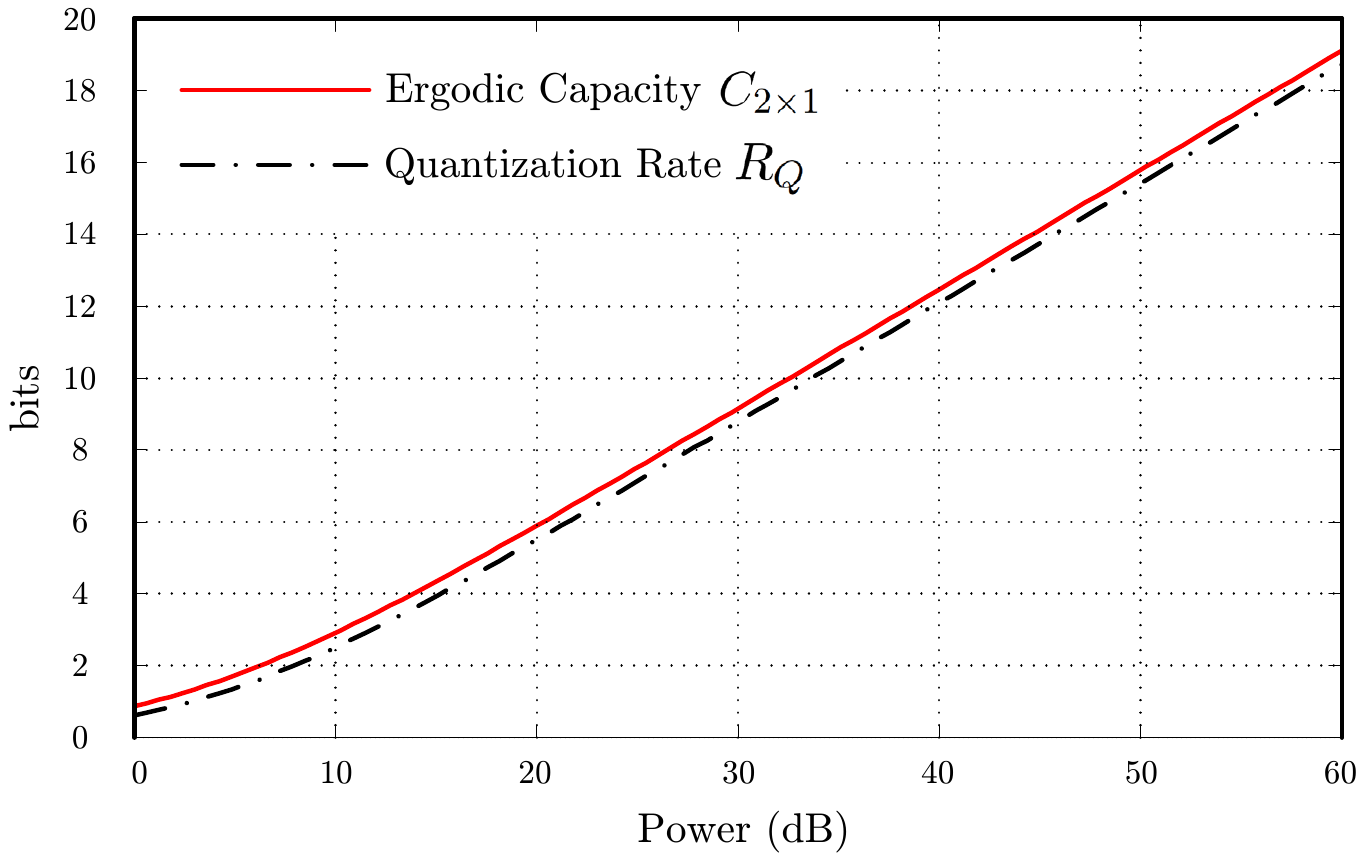}}
\caption{(a) Numerical evaluation of the per-user gap between the sum-rate inner-bound and outer-bound; (b) Using numerical analysis, we can show that $R_Q(3)/C_{2 \times 1} \leq 1$.}\label{Fig:MISO-BC-Gap}
\end{figure}

A numerical evaluation of the gap between the sum-rate inner-bound and outer-bound is plotted in Fig.~\ref{Fig:MISO-BC-Gap}(a).

To analyze the gap between the two bounds, we first study the gap between $C_{2 \times 2}(1,5)$ and $2 C_{2 \times 1}$.

\begin{corollary}
\label{CorollaryD}
Consider a MIMO point-to-point channel with $2$ transmit antennas and $2$ receive antennas as described in~\cite{Telatar-MIMO}. The only difference is that the additive noise at one antenna has variance $1$ while the additive noise at the other antenna has variance $\left( 1+D \right)$. The ergodic capacity of this channel, $C_{2 \times 2}(1,1+D)$, satisfies
\begin{align}
\label{eq:capD}
& C_{2 \times 2}(1,1+D) \nonumber \\
& \quad \geq \left( \mathbb{E}\log_2 \det \left[ \mathbf{I}_2 + \frac{P}{2} \mathbf{H}\mathbf{H}^\dagger \right] - \log_2\left( 1 + D \right) \right)^+.
\end{align}
\end{corollary}

\begin{figure*}[h!tb]
\hrule
\begin{align}
\label{eq:corollary}
& C_{2 \times 2}(1,1+D) \nonumber \\
& \overset{(a)}\geq \mathbb{E}\log_2 \det \left[ \mathbf{I}_2 + \frac{P}{2} \left[ \begin{array}{cc}
h_{11} & h_{12} \\
h_{21}/\sqrt{1+D} & h_{22}/\sqrt{1+D} \end{array} \right] \left[ \begin{array}{cc}
h_{11}^{\dagger} & h_{21}^{\dagger}/\sqrt{1+D} \\
h_{12}^{\dagger} & h_{22}^{\dagger}/\sqrt{1+D} \end{array} \right] \right] \nonumber \\
& = \mathbb{E}\log_2 \det \left[ \mathbf{I}_2 + \frac{P}{2} \left[ \begin{array}{cc}
|h_{11}|^2+|h_{12}|^2 & \left(h_{11}h_{21}^{\dagger} + h_{12}h_{22}^{\dagger} \right)/\sqrt{1+D} \\
\left(h_{11}^{\dagger}h_{21} + h_{12}^{\dagger}h_{22} \right)/\sqrt{1+D} & \left( |h_{21}|^2+|h_{22}|^2 \right)/\left(1+D\right) \end{array} \right] \right] \nonumber \\
& = \mathbb{E}\log_2 \det \left[ \begin{array}{cc}
1 + \frac{P}{2} |h_{11}|^2+|h_{12}|^2 & \frac{P}{2}\left(h_{11}h_{21}^{\dagger} + h_{12}h_{22}^{\dagger} \right)/\sqrt{1+D} \\
\frac{P}{2} \left(h_{11}^{\dagger}h_{21} + h_{12}^{\dagger}h_{22} \right)/\sqrt{1+D} & 1 + \frac{P}{2} \left( |h_{21}|^2+|h_{22}|^2 \right)/\left(1+D\right) \end{array} \right] \nonumber \\
& \geq \mathbb{E}\log_2 \det \left[ \begin{array}{cc}
1 + \frac{P}{2} |h_{11}|^2+|h_{12}|^2 & \frac{P}{2}\left(h_{11}h_{21}^{\dagger} + h_{12}h_{22}^{\dagger} \right)/\sqrt{1+D} \\
\frac{P}{2} \left(h_{11}^{\dagger}h_{21} + h_{12}^{\dagger}h_{22} \right)/\sqrt{1+D} & \left( 1 + \frac{P}{2} \left( |h_{21}|^2+|h_{22}|^2 \right) \right)/\left(1+D\right) \end{array} \right] \nonumber \\
& = \underbrace{\mathbb{E}\log_2 \det \left[ \mathbf{I}_2 + \frac{P}{2} \mathbf{H}\mathbf{H}^\dagger \right]}_{\overset{\triangle}= C_{2 \times 2}} - \log_2\left( 1 + D \right),
\end{align}
\hrule
\end{figure*}
The proof is provided at the top of the page in (\ref{eq:corollary}) where where $(a)$ holds since the right hand side is obtained by evaluating the mutual information between the input and output, for a complex Gaussian input with covariance matrix $\mathbb{E}\left[ \mathbf{x}^\dagger \mathbf{x} \right] = P/2 \mathbf{I}_2$.

Therefore, the gap between the sum-rate inner-bound and outer-bound can be upper-bounded by
\begin{align}
& \frac{4 C_{2 \times 1}}{3} - \frac{2C_{2 \times 2}(1,1+D)}{3} \nonumber \\
& \quad \leq \frac{2 \left( 2 C_{2 \times 1} - C_{2 \times 2} + \log_2 \left( 1 + D \right) \right)}{3},
\end{align}
where 
\begin{align}
% C_{2 \times 1} &\overset{\triangle}= \mathbb{E}\log_2 \left[ 1 + \frac{P}{2} \mathbf{g}^\dagger \mathbf{g} \right],% 
C_{2 \times 2} &\overset{\triangle}= \mathbb{E}\log_2 \det \left[ \mathbf{I}_2 + \frac{P}{2} \mathbf{H}\mathbf{H}^\dagger  \right].
\end{align}

\begin{remark}
While in this work we evaluate the gap for Rayleigh fading channels, our expressions for the inner-bounds and the outer-bounds hold for general i.i.d. channel realizations. The challenege to evaluate the gap for distributions other than Rayleigh fading, asises from determining the optimal covariance matrix for the transmit signal and evaluating the capacity result as discussed in~\cite{Telatar-MIMO} Section 4.1.
\end{remark}

For $P \leq 2$, the sum-rate outer-bound is smaller than $2$ bits (smaller than the gap itself). So, we assume $P>2$. We have
\begin{align}
2 & C_{2 \times 1} - C_{2 \times 2} \nonumber \\
& = 2 \mathbb{E}\log_2 \left[ 1 + \frac{P}{2} \mathbf{g}^\dagger \mathbf{g}  \right] - \mathbb{E}\log_2 \det \left[ \mathbf{I}_2 + \frac{P}{2} \mathbf{H}\mathbf{H}^\dagger  \right] \nonumber \\
& = 2 \mathbb{E}\log_2 \left[ \frac{2}{P} + \mathbf{g}^\dagger \mathbf{g}  \right] + 2 \log_2 \left( \frac{P}{2} \right) \nonumber \\
& \quad - \mathbb{E}\log_2 \det \left[ \frac{2}{P} \mathbf{I}_2 + \mathbf{H}\mathbf{H}^\dagger  \right] - \log_2 \left( \frac{P^2}{4} \right) \nonumber \\
& = 2 \mathbb{E}\log_2 \left[ \frac{2}{P} + \mathbf{g}^\dagger \mathbf{g}  \right] - \mathbb{E}\log_2 \det \left[ \frac{2}{P} \mathbf{I}_2 + \mathbf{H}\mathbf{H}^\dagger  \right] \nonumber \\
& \leq 2 \mathbb{E}\log_2 \left[ 1 + \mathbf{g}^\dagger \mathbf{g}  \right] - \mathbb{E}\log_2 \det \left[ \mathbf{H}\mathbf{H}^\dagger  \right].
\end{align}

%\begin{figure}[ht]
%\centering
%\includegraphics[height = 5cm]{FiguresPDF/C21RQforD3.pdf}
%\caption{Using numerical analysis, we can show that $R_Q(3)/C_{2 \times 1} \leq 1$.}\label{Fig:C21RQ}
%\end{figure}

Thus, we have
\begin{align}
\label{gapconst}
& \frac{4 C_{2 \times 1}}{3} - \frac{2C_{2 \times 2}(1,1+D)}{3} \nonumber \\ 
& \leq \frac{2}{3} \left( 2 \mathbb{E}\log_2 \left[ 1 + \mathbf{g}^\dagger \mathbf{g}  \right] - \mathbb{E}\log_2 \det \left[ \mathbf{H}\mathbf{H}^\dagger  \right] + \log_2 \left( 1 + D \right) \right) \nonumber \\
& \leq 3.62,
\end{align}
which implies that the gap is less than $1.81$ bits per user independent of power $P$. We could also use numerical analysis to evaluate the gap. In particular, using numerical analysis, we can show that $R_Q(3)/C_{2 \times 1} \leq 1$ (see Fig.~\ref{Fig:MISO-BC-Gap}(b)). We have plotted 
\begin{align}
\frac{4 C_{2 \times 1}}{3} - \frac{2C_{2 \times 2}(1,1+D)}{3}
\end{align}
in Fig.~\ref{Fig:MISO-BC-Gap}(a) for $D=4$, and for $P$ between $0~\mathrm{dB}$ and $60~\mathrm{dB}$. As we can see, the sum-rate inner-bound and outer-bound are at most $2.2$ bits (or $1.1$ per user) away from each other for $P$ between $0~\mathrm{dB}$ and $60~\mathrm{dB}$.

\section{Symmetric Capacity of the $K$-user MISO BC with Delayed CSIT}
\label{Section:Kuser}

\begin{figure}[t]
\centering
\includegraphics[height = 5cm]{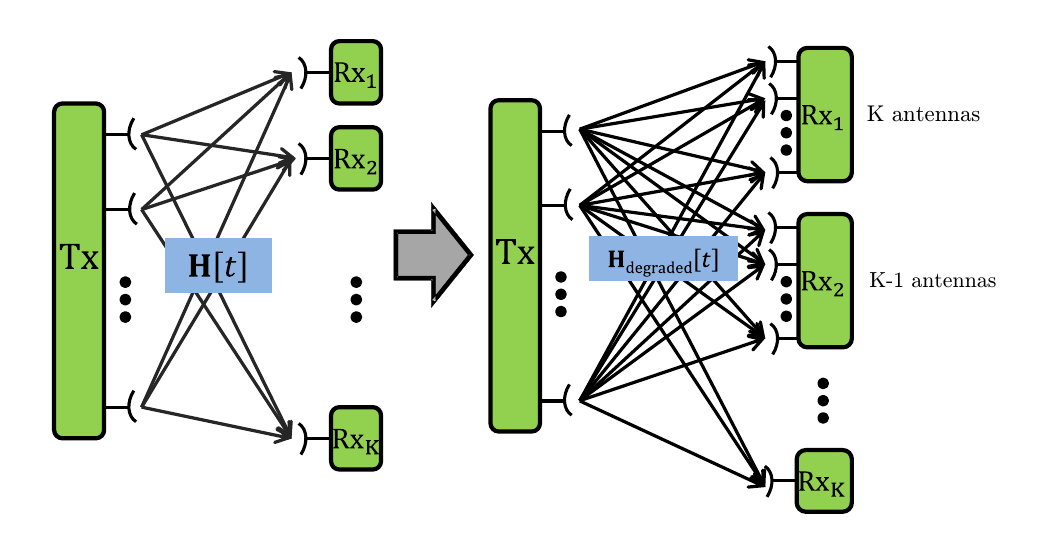}
\caption{Left: $K$-user Multiple-Input Single-Output (MISO) Complex Gaussian Broadcast Channel; and right: degraded MIMO BC.}\label{Fig:MISO-BC-K-User}
\end{figure}

Now that we have presented our results for the two-user multiple-input single-output complex Gaussian broadcast channel with delayed CSIT, we consider the $K$-user setting as depicted in Fig.~\ref{Fig:MISO-BC-K-User} (left). The channel matrix from the transmitter to the receivers is denoted by $\mathbf{H} \in \mathbb{C}^{K \times K}$, where the entries of $\mathbf{H}[t]$ are distributed as i.i.d. $\mathcal{CN}(0,1)$ (independent across time, antenna, and users). The transmit signal $\mathbf{x}[t] \in \mathbb{C}^{K \times 1}$ is subject to average power constraint $P$, \emph{i.e.} $\mathbb{E}\left[ \mathbf{x}^\dagger[t] \mathbf{x}[t] \right] \leq P$ for $P > 0$. The noise processes are independent from the transmit signal and are distributed i.i.d. as $z_k[t] \sim \mathcal{CN}(0,1)$. The input-output relationship in this channel is given by
%\begin{align}
%\left[ \begin{array}{c}
%y_{1}[t] \\
%y_{2}[t] \\
%\vdots   \\
%y_{K}[t] \end{array} \right] =
%\mathbf{H}[t] \mathbf{x}[t] + 
%\left[ \begin{array}{c}
%z_{1}[t] \\
%z_{2}[t] \\
%\vdots   \\
%z_{K}[t] \end{array} \right].
%\end{align}
\begin{align}
\left[~y_{1}[t] \quad \ldots \quad y_{K}[t] \right]^\top = \mathbf{H}[t] \mathbf{x}[t] + \left[~z_{1}[t] \quad \ldots \quad z_{K}[t]~\right].
\end{align}

\subsection{Outer-bound}
\label{Section:KUserOuter}

The derivation of the outer-bound in Theorem~\ref{THM:Kuser} is based on creating a degraded MIMO BC where ${\sf Rx}_{\ell}$ has access to the received signal of ${\sf Rx}_j$ for $j \ge \ell$ similar to the converse proof provided in~\cite{MAT_Delayed}, $\ell=1,2,\ldots,K$. This channel is physically degraded, and thus feedback does not enlarge the capacity region~\cite{ElGamal-Degraded}. Therefore, we can ignore the delayed knowledge of the channel state information at the transmitter (\emph{i.e.} no CSIT assumption) and the correlation between the channel gains. See Fig.~\ref{Fig:MISO-BC-K-User} for a depiction. In this degraded channel, we denote the output signal of user $j$ at time $t$ by $\tilde{\mathbf{y}}_j$\footnote{for user $K$, we have $\tilde{\mathbf{y}}_K[t] = \tilde{y}_K[t] = y_K[t]$} where
\begin{align}
\tilde{\mathbf{y}}_j[t] = 
\left[~y_{j}[t]~y_{j+1}[t]~\ldots~y_{K}[t]~\right]^\top.
\end{align}

For the resulting multiple-input multiple-output (MIMO) broadcast channel denote the channel matrix by $\mathbf{H}_{\mathrm{degraded}}^n$. The capacity region of the $K$-user MISO BC with delayed CSIT (Fig.~\ref{Fig:MISO-BC-K-User} left) is included in the capacity region of the degraded MIMO BC with no CSIT (Fig.~\ref{Fig:MISO-BC-K-User} right). Under no CSIT assumption, the following result is known for the stochastically degraded MIMO BC described above. 

\begin{lemma}[\hspace{-.1mm}\cite{vaze2012degree}]
\label{Lemma:MIMOBC}
Consider the $K$-user stochastically degraded MIMO BC described above and depicted in Fig.~\ref{Fig:MISO-BC-K-User} (right). Then under no CSIT assumption, for $j=1,\ldots,K-1,$ we have
\begin{align}
h &\left( \tilde{\mathbf{y}}_{j}^n | w_{j+1}, \ldots, w_{K}, \mathbf{H}_{\mathrm{degraded}}^n \right) \nonumber \\
& \geq \frac{K-j+1}{K-j} h\left( \tilde{\mathbf{y}}_{j+1}^n | w_{j+1}, \ldots, w_{K}, \mathbf{H}_{\mathrm{degraded}}^n \right). 
\end{align}
\end{lemma} 

Using Lemma~\ref{Lemma:MIMOBC}, similar to the argument presented in (\ref{eq:boundR12R2}), we get
\begin{align}
n & \left( R_1 + \frac{K}{K-1} R_2 + \frac{K}{K-2} R_2 + \ldots + K R_K - \epsilon_n \right) \nonumber \\
& = I\left( w_1; \tilde{\mathbf{y}}_1^n | w_{2}, \ldots, w_{K}, \mathbf{H}_{\mathrm{degraded}}^n \right) \nonumber
\end{align}
\begin{align}
& + \frac{K}{K-1} I\left( w_2; \tilde{\mathbf{y}}_2^n | w_{3}, \ldots, w_{K}, \mathbf{H}_{\mathrm{degraded}}^n \right) \nonumber \\
& + \ldots + K I\left( w_K; \tilde{y}_K^n | \mathbf{H}_{\mathrm{degraded}}^n \right) \nonumber \\
& \leq h\left(\tilde{\mathbf{y}}_1^n | w_{2}, \ldots, w_{K}, \mathbf{H}_{\mathrm{degraded}}^n \right) \nonumber \\
& - \frac{K}{K-1} h\left(\tilde{\mathbf{y}}_2^n | w_{2}, \ldots, w_{K}, \mathbf{H}_{\mathrm{degraded}}^n \right) \nonumber \\
& + \frac{K}{K-1} h\left(\tilde{\mathbf{y}}_2^n | w_{3}, \ldots, w_{K}, \mathbf{H}_{\mathrm{degraded}}^n \right) \nonumber \\
& - \frac{K}{K-2} h\left(\tilde{\mathbf{y}}_3^n | w_{3}, \ldots, w_{K}, \mathbf{H}_{\mathrm{degraded}}^n \right) \nonumber \\
& + \ldots + K h\left(\tilde{y}_K^n | \mathbf{H}_{\mathrm{degraded}}^n \right) \nonumber \\
& \overset{Lemma~\ref{Lemma:MIMOBC}}\leq K h\left(\tilde{y}_K^n | \mathbf{H}_{\mathrm{degraded}}^n \right) \nonumber \\
& \leq K \mathbb{E}\log_2 \left[ 1 + \frac{P}{K} \mathbf{g}^\dagger \mathbf{g} \right] \overset{(\ref{Def:Ckl})}= K C_{K \times 1},
\end{align}
where $\mathbf{g}$ is a $K$ by $1$ vector where entries are i.i.d. $\mathcal{CN}\left( 0, 1\right)$. Then, for the symmetric sum-rate point we have $R_1 = R_2 = \ldots = R_K = R$, which gives us
\begin{align}
\left( \sum_{j=1}^{K}{j^{-1}} \right) R \leq K C_{K \times 1},
\end{align}
which implies that
\begin{align}
C_{\mathrm{SYM},\mathrm{K-user}} \leq \frac{C_{K \times 1}}{\sum_{j=1}^{K}{j^{-1}}}.
\end{align}
This completes the converse proof of Theorem~\ref{THM:Kuser}.

\subsection{Achievability}

In this subsection, we focus on achievability and discuss the extension of our achievability results to the $K$-user MISO BC with delayed CSIT. The transmission strategy follows the steps of~\cite{MAT_Delayed}, but similar to the two-user case, some additional ingredients are needed to make an approximately optimal scheme. We first demonstrate the techniques and the required ingredients for the three-user MISO BC with delayed CSIT. Then, we discuss how the result is extended to the general case.

{\bf Transmission strategy for the three-user MISO BC}: Each communication block includes three phases where Phase~1 and Phase~2 each have $3$ segments, and Phase~3 has two segments. We focus on a specific block $b$. Fix $\epsilon > 0$.

$\bullet$ During segments $1,2,$ and $3$ of Phase $1$ of block $b$, the message of user $k$ ($w^b_k$) is encoded as $\mathbf{x}_k^{b,n}$ at rate
\begin{align}
\label{eq:3userrateC}
C_{3 \times 3}\left(1, 5, 11 \right) - \epsilon,
\end{align}
where any element of this codeword is drawn i.i.d. from $\mathcal{CN} \left( 0, P/3 \mathbf{I}_3 \right)$\footnote{$\mathbf{I}_3$ is the $3 \times 3$ identity matrix.}; and $C_{3 \times 3}\left( 1, 5, 11 \right)$ is given in Definition~\ref{Def:Ckl}. 

\begin{remark}
The reason for the encoding rate of (\ref{eq:3userrateC}) becomes apparent as we describe the achievability strategy. Basically the quantization noise accumulates on top of the previous noises throughout some of the phases and thus, we need to adjust the encoding rate accordingly.
\end{remark}

Transmitter communicates these codewords and receiver one obtains $y_{1,1:j}^{b,n}$, receiver two obtains $y_{2,1:j}^{b,n}$, and receiver three obtains $y_{3,1:j}^{b,n}$, $j=1,2,3$. 

\begin{definition}
We define $s_{k,p:j}^{b,n}$ to be the noiseless version of $y_{k,p:j}^{b,n}$ for appropriate choice of indices. This is similar to Definition~\ref{defineS} for the two-user case.
\end{definition}

Note that at the end of the third segment of the first phase, transmitter has access to $s_{1,1:j}^{b,n}$, $s_{2,1:j}^{b,n}$, and $s_{3,1:j}^{b,n}$. Due to the rate given in (\ref{eq:3userrateC}), if $s_{2,1:1}^{b,n}$ and $s_{3,1:1}^{b,n}$ are provided to receiver one with distortions $4$ and $8$ respectively, then receiver one will be able to decode its message with arbitrary small decoding error probability. Similarly receiver two is interested in $s_{1,1:2}^{b,n}$ and $s_{3,1:2}^{b,n}$; and receiver three is interested in $s_{1,1:3}^{b,n}$ and $s_{2,1:3}^{b,n}$.

Based on the discussion above and the available signal at each receiver, we observe that $s_{2,1:1}^{b,n} + s_{1,1:2}^{b,n}$ is of common interest to receivers one and two. Similarly, $s_{3,1:1}^{b,n} + s_{1,1:3}^{b,n}$ is of common interest to receivers one and three; and $s_{2,1:3}^{b,n} + s_{3,1:2}^{b,n}$ is of common interest to receivers two and three. Therefore, the goal would be to deliver these signals to their interested receivers during the following phases. 

$\bullet$ Communication during Phase~$2$ Segment~$1$: Consider the communication during another block $b^\prime$, and the corresponding signal
$s_{2,1:1}^{b^\prime,n} + s_{1,1:2}^{b^\prime,n}$. Using Lemma~\ref{Lemma:RateDistortion}, we quantize $s_{2,1:1}^{b,n} + s_{1,1:2}^{b,n}$ and $s_{2,1:1}^{b^\prime,n} + s_{1,1:2}^{b^\prime,n}$ at distortion $4$, and we create the XOR of the resulting bits\footnote{We note that we need these signals to be distributed independently, to handle this issue, we can incorporate an interleaving step similar to Section~\ref{Section:Achievability}.}. Then these bits will be encoded as $\mathbf{v}_{2:1}^{b,n}$ at rate $C_{3 \times 2}\left(1, 5 \right) - \epsilon$, and will be transmitted during the first segment of Phase~$2$ where $C_{3 \times 2}\left( 1, 5 \right)$ is given in Definition~\ref{Def:Ckl}. Receiver three obtains $y_{3,2:1}^{b,n}$ that is of interest of users one and two.

$\bullet$ Communication during Phase~$2$ Segment~$2$: Consider the communication during another block $b^\prime$, and the corresponding signal
$s_{3,1:1}^{b^\prime,n} + s_{1,1:3}^{b^\prime,n}$. Using Lemma~\ref{Lemma:RateDistortion}, we quantize $s_{3,1:1}^{b,n} + s_{1,1:3}^{b,n}$ and $s_{3,1:1}^{b^\prime,n} + s_{1,1:3}^{b^\prime,n}$ at distortion $4$ and we create the XOR of the resulting bits. Then these bits will be encoded at rate $C_{3 \times 2}\left(1, 5 \right) - \epsilon$ denoted by $\mathbf{v}_{2:2}^{b,n}$ and will be transmitted during the second segment of Phase~$2$. Receiver two obtains $y_{2,2:2}^{b,n}$ that is of interest of users one and three.

$\bullet$ Communication during Phase~$2$ Segment~$3$: Consider the communication during another block $b^\prime$, and the corresponding signal
$s_{2,1:3}^{b^\prime,n} + s_{3,1:2}^{b^\prime,n}$. Using Lemma~\ref{Lemma:RateDistortion}, we quantize $s_{2,1:3}^{b,n} + s_{3,1:2}^{b,n}$ and $s_{2,1:3}^{b^\prime,n} + s_{3,1:2}^{b^\prime,n}$ at distortion $4$ and we create the XOR of the resulting bits. Then these bits will be encoded at rate $C_{3 \times 2}\left(1, 5 \right) - \epsilon$ denoted by $\mathbf{v}_{2:3}^{b,n}$ and will be transmitted during the third segment of Phase~$3$. Receiver one obtains $y_{1,2:3}^{b,n}$ that is of interest of users two and three.

We now create two signals that are of interest of all three receivers: 
\begin{align}
\label{eq:order3}
% s_{3,4}^{b,n} + \frac{\sqrt{2}}{\sqrt{3}} s_{2,5}^{b,n} + \frac{1}{\sqrt{3}} s_{1,6}^{b,n}, \qquad \text{and} \qquad \frac{1}{\sqrt{3}} s_{3,4}^{b,n} - \frac{\sqrt{2}}{\sqrt{3}} s_{2,5}^{b,n} +  s_{1,6}^{b,n}.
& \frac{1}{\sqrt{2}} s_{3,2:1}^{b,n} + \frac{1}{\sqrt{3}} s_{2,2:2}^{b,n} + \frac{1}{\sqrt{6}} s_{1,2:3}^{b,n}, \nonumber \\
& \frac{1}{\sqrt{6}} s_{3,2:1}^{b,n} + \frac{-1}{\sqrt{3}} s_{2,2:2}^{b,n} + \frac{1}{\sqrt{2}} s_{1,2:3}^{b,n}.
\end{align}

\begin{remark}
The choice of coefficients in (\ref{eq:order3}) is such that all users are assigned equal powers and the linear combinations at the receivers remain independent. We also note that using such coefficients results in a $1/3$ power loss for each user.
\end{remark}

Note that any receiver has access to both signals in (\ref{eq:order3}), then it can recursively recover the signals it is interested in, and decode the intended message.

$\bullet$ Communication during Phase~$3$ Segment~$1$: Using Lemma~\ref{Lemma:RateDistortion}, we quantize $$\left( \frac{1}{\sqrt{2}} s_{3,2:1}^{b,n} + \frac{1}{\sqrt{3}} s_{2,2:2}^{b,n} + \frac{1}{\sqrt{6}} s_{1,2:3}^{b,n} \right)$$ at distortion $5$. Then these quantized bits will be encoded at rate $C_{3 \times 1} - \epsilon$ and transmitted during the first segment of Phase~$3$.

$\bullet$ Communication during Phase~$3$ Segment~$2$: Using Lemma~\ref{Lemma:RateDistortion}, we quantize $$\left( \frac{1}{\sqrt{6}} s_{3,2:1}^{b,n} + \frac{-1}{\sqrt{3}} s_{2,2:2}^{b,n} + \frac{1}{\sqrt{2}} s_{1,2:3}^{b,n} \right)$$ at distortion $5$. Then these quantized bits will be encoded at rate $C_{3 \times 1} - \epsilon$ and transmitted during the second segment of Phase~$3$.
 
Using the achievability strategy described above and for $n \rightarrow \infty$ and $\epsilon \rightarrow 0$, it can be shown that a per-user rate of 
% \begin{align}
$\frac{2}{11}C_{3 \times 3}\left( 1, 5, 11 \right)$
% \end{align}
is achievable.

{\bf Transmission strategy for the $K$-user MISO BC}: Now that we have described the transmission strategy for the $3$-user MISO BC with delayed CSIT, we explain the transmission strategy for the general case $K > 3$. As mentioned before, the transmission strategy follows the steps of~\cite{MAT_Delayed}. However, we highlight the key differences that are needed to derive near optimal results.

The scheme includes $K$ phases. For simplicity, we do not go into details of the interleaving process. At the beginning of Phase $j$, transmitter has access to signals that are of interest of $j$ receivers, $j=1,2,\ldots,K$. There are a total of $\left( K - j + 1 \right) {K \choose j}$ such signals. Phase $j$ has ${K \choose j}$ segments. Consider a subset $\mathcal{S}$ of the receivers where $\left| \mathcal{S} \right| = j$. Transmitter accumulates a total of $\left( K - j + 1 \right)$ signals that are of interest of all receivers in $\mathcal{S}$ using $\left( K - j + 1 \right)$ different blocks (this is similar to Phase~2 for the $3$-user MISO BC with delayed CSIT). Similar to Lemma~\ref{Lemma:RateDistortion}, transmitter quantizes these signals at distortion $\left( j + 2 \right)$, and creates the XOR of the resulting bits. The resulting bits are encoded at rate $C_{K \times j}\left(1, 5,\ldots, 1+\left( j - 1 \right)\left( j + 2 \right) \right) - \epsilon$ and communicated during segment $t_{\mathcal{S}}$ of Phase~$j$. 

\begin{remark}
The noise variance $1 + \left( j - 1 \right)\left( j + 2 \right)$ results from the fact that each receiver has to solve $\left( j - 1 \right)$ equations of the signals that he is interested in. This step results in boosting up the noise variance.
\end{remark}

Consider any subset $\mathcal{S}^\prime$ of the receivers where $\left| \mathcal{S}^\prime \right| = j+1$. Upon completion of Phase~$j$, we observe that any receiver in $\mathcal{S}^\prime$ has a signal that is simultaneously of common interest of all other receivers in $\mathcal{S}^\prime$. Transmitter has access to this signal (up to noise) using delayed knowledge of the channel state information. Transmitter forms $j$ random linear combinations of such signals for each subset $\mathcal{S}^\prime$ where $\left| \mathcal{S}^\prime \right| = j+1$. Then transmitter creates  $j {K \choose j+1}$ signals that are simultaneously of interest of $j+1$ receivers. These signals (after being quantized) will be delivered in Phases~$j+1,j+2,\ldots,K$. The rest of the scheme is identical to that of~\cite{MAT_Delayed} and is omitted due to space limitations. 

Recursively solving the achievable rate over $K$ phases, we can show that a per-user rate of 
{\small \begin{align}
\frac{C_{K \times K}\left( 1, 5, \ldots, 1+(j-1)(j+2), \ldots, 1+(K-1)(K+2) \right)}{K\sum_{j=1}^{K}{j^{-1}}}
\end{align}}
is achievable. 

We are now ready to present the approximate capacity of the $K$-user MISO BC with delayed CSIT.

\begin{corollary}
\label{Corollary:CapacityKuser}
The capacity region of the $K$-user MIMO BC with delayed CSIT is within at most $2\log_2(K+2)$ bits per user of the region described by:
\begin{align}
\label{eq:JBounds}
0 \leq & R_{\pi_1} + \frac{\left| \mathcal{J} \right|}{\left| \mathcal{J} \right| -1} R_{\pi_2} + \frac{\left| \mathcal{J} \right|}{\left| \mathcal{J} \right| -2} R_{\pi_3} \\
& + \ldots +  \left| \mathcal{J} \right| R_{\pi_{\left| \mathcal{J} \right|}} \leq \left| \mathcal{J} \right| C_{K \times 1}, \quad \forall \mathcal{J} \subseteq \{ 1,2,\ldots,K \}, \nonumber
\end{align}
for all permutations $\pi$ of $\mathcal{J}$.
\end{corollary}

The proof is based on the results we provided in this section and thus, we only provide a proof sketch here. The derivation of the bounds is similar to that of Section~\ref{Section:KUserOuter} for a degraded MIMO BC with $K$ transmit antennas and $\left| \mathcal{J} \right|$ receivers. 

To obtain the achievability, we use the transmission strategy that we described in this section with a small modification. To explain why a modification is needed, consider $\left| \mathcal{J} \right| = 2$. Then we obtain ${K \choose 2}$ pair of bounds similar to:
\begin{align}
R_1 + 2 R_2 \leq 2 C_{K \times 1}, \nonumber\\
2 R_1 + R_2 \leq 2 C_{K \times 1}. 
\end{align}
These bounds differ from the bounds in Sections~\ref{Section:Achievability} and~\ref{Section:Converse} since they have  $C_{K \times 1}$ rather than $C_{2 \times 1}$on the right hand side. Thus our transmission strategy needs to be modified to capture this difference.

Fix $\mathcal{J}$, then from (\ref{eq:JBounds}) we obtain $\left| \mathcal{J} \right|!$ bounds that will define a symmetric sum-rate point (symmetric for the users in $\mathcal{J}$) of
\begin{equation}
\label{eq:JCornerPoint}
R_i =
\left\{ \begin{array}{ll}
\vspace{1mm} 0, & \text{if~} i \notin \mathcal{J}, \\
\left( {\sum_{j=1}^{\left| \mathcal{J} \right|}{j^{-1}}} \right)^{-1}C_{K \times 1}, & \text{if~} i \in \mathcal{J}.
\end{array} \right.
\end{equation}
It is sufficient to show we can achieve to within a constant gap of all such corner points. 

In this section, we presented the transmission strategy of the symmetric sum-rate point of $K$-user MISO BC with delayed CSIT. It is easy to modify this result to achieve within a constant gap of the corner point in (\ref{eq:JCornerPoint}). Start with a  $\left| \mathcal{J} \right|$-user MISO BC and re-write the transmission strategy of this section for that problem. Then, we need to modify the scheme by replacing $C_{\left| \mathcal{J} \right| \times \ell}$ whenever it appears with $C_{K \times \ell}$, $1 \leq \ell \leq \left| \mathcal{J} \right| \leq K$ and the result would follow. Finally, we note that using an argument similar to the one presented in Section~\ref{Section:Gap}, the gap takes its maximum value at the symmetric sum-rate point of the $K$-user setting. Thus, the gap is at most $2\log_2(K+2)$ bits per user.

\subsection{Gap Analysis}

\begin{figure}[ht]
\centering
\includegraphics[height = 6cm]{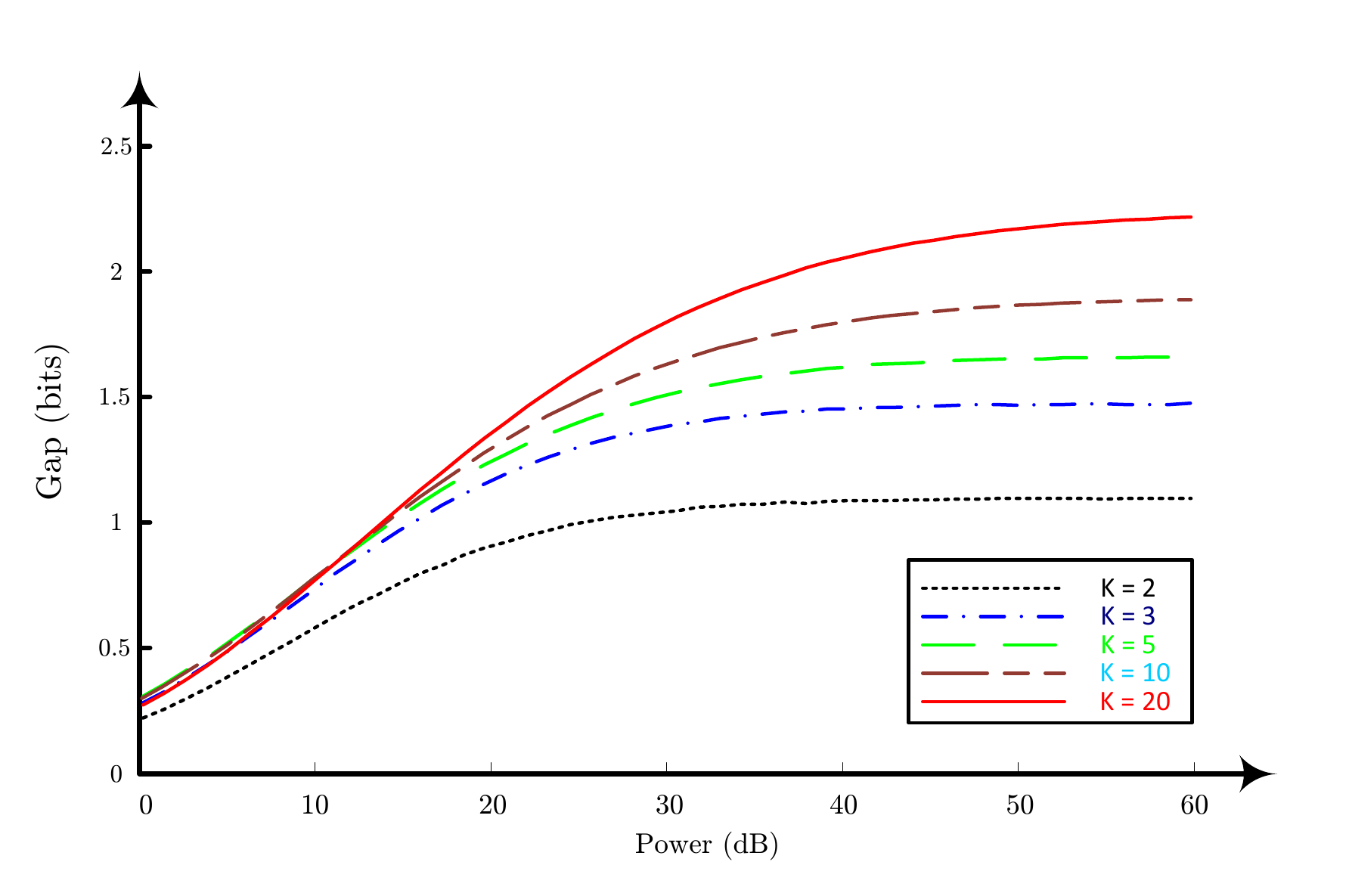}
\caption{Numerical evaluation of the per user gap between the inner-bound and the outer-bound in Theorem~\ref{THM:Kuser} for $P$ between $0~\mathrm{dB}$ and $60~\mathrm{dB}$ and $K=2,3,5,10,20$.}\label{Fig:MISO-BC-Gap-Kuser}
\end{figure}

In Fig.~\ref{Fig:MISO-BC-Gap-Kuser}, we have plotted the numerical analysis of the per user gap between the inner-bound and the outer-bound of Theorem~\ref{THM:Kuser} for $P$ between $0~\mathrm{dB}$ and $60~\mathrm{dB}$ and $K=2,3,5,10,20$.

For the per user gap, we analytically show that the outer-bound and lower-bound given in Theorem~\ref{THM:Kuser} are at most $2\log_2(K+2)$ away from each other. However as depicted in Fig.~\ref{Fig:MISO-BC-Gap-Kuser}, we can see that the per-user gap is in fact smaller than $2\log_2(K+2)$. The argument is similar to the one we presented in (\ref{eq:corollary}). To derive the gap, we first increase the noise term at all receive antennas to $1+(K-1)(K+2)$. Then similar to (\ref{eq:corollary}), we obtain
{\small \begin{align}
\label{eq:gapKuser}
& C_{K \times K}\left( 1, 5, \ldots, 1+(j-1)(j+2), \ldots, 1+(K-1)(K+2) \right) \nonumber \\
& \ge C_{K \times K} - K \log_2\left( 1 + (K-1)(K+2) \right).
\end{align}}

Using (\ref{eq:gapKuser}) and considering the gap between $C_{K \times K}$ and $KC_{K \times 1}$, we can show that the gap between the achievable symmetric rate and the symmetric capacity is bounded by $2\log_2(K+2)$.

%%%%%%%%%%%%%%%%%%%%%%%%%%%%%%%%%%%%%%%%%%%%%%%%%%%
%%%%%%%%%%%%%%%%%%%%%%%%%%%%%%%%%%%%%%%%%%%%%%%%%%%
%%%%%%%%%%%%%%%%%%%%%%%%%%%%%%%%%%%%%%%%%%%%%%%%%%%

\section{Conclusion and Future Directions}
\label{Section:Conclusion}

In this paper, we studied the capacity region of the multiple-input single-output complex Gaussian Broadcast Channels with delayed CSIT. We showed that a modification of the scheme introduced in~\cite{MAT_Delayed}, can be applied in the finite SNR regime to obtain an inner-bound that is within $2\log_2(K+2)$ bits of the outer-bound. Therefore the gap scales as the logarithm of the number of users. This happens due to noise accumulation during the transmission strategy. Thus, an interesting future direction would be to figure out whether there exists a transmission strategy that results in constant gap (independent of channel parameters, transmission power and number of users) approximation of the capacity region. 

Another direction is to consider a two-user MISO BC with delayed CSIT where the noise processes and the channel gains are not distributed as i.i.d. random variables. For example, consider the scenario where the noise processes have different variances. This model captures the scenario where users are located at different distances to the base station. For this setting, even the (generalized) DoF region is not known.

\appendices
%
%\section{Lattice Quantizer}
%\label{Appendix:Lattice}
%
%\input{PreliminariesLattice.tex}
%
%
\section{Determining $D$ such that $R_Q(D)/C_{2 \times 1} \leq 1$}
\label{Appendix:ValD}

As mentioned in Section~\ref{Section:Gap}, we are interested in $P>2$. Using Jensen's inequality, we have
\begin{align} 
& R_Q(D) = \mathbb{E}\left[ \log_2 \left( 1 + \frac{P}{2D} \left( ||\mathbf{g}||_2^2 + ||\mathbf{h}||_2^2 \right) \right)\right] \\
& \leq \log_2 \left( 1 + \frac{P}{2D} \mathbb{E}\left[ ||\mathbf{g}||_2^2 + ||\mathbf{h}||_2^2 \right] \right) = \log_2 \left( 1 + \frac{2P}{D} \right). \nonumber 
\end{align}

Moreover, from~\cite{Telatar-MIMO}, we have
\begin{align}
& C_{2 \times 1} = \int_{0}^{\infty}{\log_2\left(1 + P\lambda/2 \right) \lambda e^{-\lambda}d\lambda} \nonumber \\
& = \int_{0}^{1}{\log_2\left(1 + P\lambda/2 \right) \lambda e^{-\lambda}d\lambda} \nonumber \\
& + \int_{1}^{\infty}{\log_2\left(1 + P\lambda/2 \right) \lambda e^{-\lambda}d\lambda} \nonumber \\
& = \sum_{m=1}^{\infty}{\int_{2^{-m}}^{2^{1-m}}{\log_2\left(1 + P\lambda/2 \right) \lambda e^{-\lambda}d\lambda}}  \nonumber \\
& + \int_{1}^{2}{\log_2\left(1 + P\lambda/2 \right) \lambda e^{-\lambda}d\lambda} \nonumber \\
& + \sum_{j=1}^{45}{\int_{2+.1(j-1)}^{2+.1j}{\log_2\left(1 + P\lambda/2 \right) \lambda e^{-\lambda}d\lambda}} \nonumber \\
& + \int_{6.5}^{\infty}{\log_2\left(1 + P\lambda/2 \right) \lambda e^{-\lambda}d\lambda} \nonumber \\
& \overset{(a)}\geq \log_2\left(1 + P/2 \right) \int_{0}^{1}{\lambda e^{-\lambda}d\lambda} - \underbrace{\sum_{m=1}^{\infty}{m\int_{2^{-m}}^{2^{1-m}}{\lambda e^{-\lambda}d\lambda}}}_{< 0.4} \nonumber \\
& + \log_2\left(1 + P/2 \right) \int_{1}^{2}{\lambda e^{-\lambda}d\lambda} \nonumber \\
&  + \log_2\left(1 + P/2 \right) \int_{2}^{6.5}{\lambda e^{-\lambda}d\lambda} \nonumber \\
& + \underbrace{\sum_{j=1}^{45}{\log_2\left[ \frac{1+(2+.1(j-1))P/2}{1+P/2} \right]\int_{2+.1(j-1)}^{2+.1j}{\lambda e^{-\lambda}d\lambda}}}_{> 0.4} \nonumber \\
&  + \int_{6.5}^{\infty}{\log_2\left(1 + P\lambda/2 \right) \lambda e^{-\lambda}d\lambda} \nonumber \\
& > \log_2\left(1 + P/2 \right) \int_{0}^{\infty}{\lambda e^{-\lambda}d\lambda} = \log_2\left(1 + P/2 \right).
\end{align}
where $(a)$ holds since
\begin{align}
& \sum_{m=1}^{\infty}{\int_{2^{-m}}^{2^{1-m}}{\log_2\left(1 + P\lambda/2 \right) \lambda e^{-\lambda}d\lambda}} \nonumber \\
& \geq \sum_{m=1}^{\infty}{\int_{2^{-m}}^{2^{1-m}}{\log_2\left(1 + 2^{-m}P/2 \right) \lambda e^{-\lambda}d\lambda}} \\
& \geq \sum_{m=1}^{\infty}{\int_{2^{-m}}^{2^{1-m}}{\left[ \log_2\left(1 + P/2 \right) - \log_2\left( 2^m \right) \right] \lambda e^{-\lambda}d\lambda}} \nonumber \\
& = \log_2\left(1 + P/2 \right) \int_{0}^{1}{\lambda e^{-\lambda}d\lambda} - \sum_{m=1}^{\infty}{m\int_{2^{-m}}^{2^{1-m}}{\lambda e^{-\lambda}d\lambda}}, \nonumber 
\end{align}
and $\sum_{m=1}^{\infty}{m\int_{2^{-m}}^{2^{1-m}}{\lambda e^{-\lambda}d\lambda}}$ converges since
\begin{align}
\left\{ m\int_{2^{-m}}^{2^{1-m}}{\lambda e^{-\lambda}d\lambda} \right\}_{m=1}^{\infty},
\end{align}
is a Cauchy sequence. Thus, we have $R_Q(4)/C_{2 \times 1} < 1$.

%\section{Gap Analysis}
%\label{Appendix:Gap}
%
%\input{AppGap.tex}

\bibliographystyle{ieeetr}
\bibliography{bib_misobc}

\end{document}